\documentclass[12pt,preprint]{aastex63}

\providecommand{\mum}{\ifmmode{\rm \mu m}\else{$\mu$m}\fi}

\begin{document}

\newcommand{\mjysr}{MJy sr$^{-1}$}

\title{SMC-Last Mosaic Images}

\begin{abstract}
We present mosaic images of the Small Magellanic Cloud (SMC) observed with
the Spitzer IRAC 3.6 $\micron$ and 4.5 $\micron$ bands over two epochs,
2017 August 25 to 2017 September 13, and 2017 November 24 to 2018 February 12. 
The survey region comprises $\sim$30 square degrees covering
the SMC and the Bridge to the Large Magellanic Cloud. The region is
covered by 52 $\sim$1$\fdg$1$\times$1$\fdg$1 tiles, with each tile including images
in each band for both 
separate and combined epochs. The mosaics are made in individual tangent projections
in J2000 coordinates. The angular pixel size is 0${\farcs}$6 with a
resolution (FWHM) of $\sim$2${\farcs}$0. We describe
processing to correct or mitigate residual artifacts and remove
background discontinuities.
The mosaic images are publicly available at the Infrared Science 
Archive (IRSA).
\end{abstract}

\author[0000-0003-0947-2824]{D. R. Mizuno}
\affiliation{Institute for Scientific Research, Boston College,\\
140 Commonwealth Avenue, Chestnut Hill, MA 02467, USA}

\author[0000-0002-2626-7155]{Kathleen E. Kraemer}
\affiliation{Institute for Scientific Research, Boston College,\\
140 Commonwealth Avenue, Chestnut Hill, MA 02467, USA}

\author[0000-0003-1955-8509]{T. A. Kuchar}
\affiliation{Institute for Scientific Research, Boston College,\\
140 Commonwealth Avenue, Chestnut Hill, MA 02467, USA}

\author[0000-0003-4520-1044]{G. C. Sloan}
\affiliation{Space Telescope Science Institute, 3700 San Martin
Drive, Baltimore, MD 21218, USA}
\affiliation{Department of Physics and Astronomy, University of
  North Carolina, Chapel Hill, NC 27599-3255, USA}

\section{Introduction}

The Small Magellanic Cloud (SMC) is a nearby metal-poor dwarf galaxy.
Its nominal distance is 62 kpc \citep[e.g.,][]{deg15,gra20}, but 
its structure is complex, with a foreground component at a distance
of $\sim$55 kpc from the Sun on the east side, likely due to tidal
interactions with the Large Magellanic Cloud (LMC) \citep{nid13,sub17} and another component
several kpc behind the southwest regions of the galaxy 
\citep[e.g.,][]{sco16,yan21}.  The metallicity of the SMC depends on
the sample observed.  Cepheids give [Fe/H] = $-$0.68 \citep{luc98},
while for the red-giant population, the mean [Fe/H] = $-$0.97 
\citep{cho20}.  \cite{rub18} find that [Fe/H] ranges from
$\sim$$-$0.6 for the youngest population to $\sim$$-$1.6 for the
oldest.

Thus the SMC serves, on the one hand, as a laboratory for studying
the evolution of a population of stars as a whole, at a known distance and at
a metallicity significantly lower than the Galaxy.  On the
other hand, the SMC is a complex and dynamic metal-poor galaxy
close enough to our own that it can be studied in great detail.  

Understanding the chemical evolution of this galaxy requires 
observations of the stars both as they die, when they enrich the 
SMC with freshly fused elements and dust, and as they form.
Because stars evolving to and from the main sequence are variables,  
a proper study of those populations requires multi-epoch surveys.  
The Optical Gravitational Lensing Experiment \citep[OGLE; 
e.g.,][]{uda97,uda08} has monitored the SMC at a near-daily cadence 
from 1997.  The OGLE-III catalogs include data through 2008 and have 
identified over 19,000 long-period variables in the SMC, including 
over 2,000 variables likely on the asymptotic giant branch 
\citep{sos11}.

Photometry in the near-infrared is more sensitive to the cooler
objects evolving to and from the main sequence.  The VISTA Magellanic
Cloud Survey (VMC) surveyed 45 square degrees in the SMC at $Y$, $J$, 
and $K_s$, with 12 epochs at $K_s$, starting in 2011 \citep{cio11}.  
The SMC region of the VMC covers the Bar, which includes most of the 
well-known \ion{H}{2} regions in the galaxy, and the Wing, which 
extends to the east, toward the LMC.  An additional 20 square degrees 
covers part of the Bridge between the Magellanic Clouds.  The Infrared 
Survey Facility mapped the central square degree at $J$, $H$, and $K_s$ 
\citep{ita18}.  While not as deep as the VMC, it has better temporal 
coverage, with over 100 epochs from 2001 to 2017.

The mid-infrared is best suited to study the most embedded objects, 
the stars that are actually forming or dying.  Using the Spitzer
Space Telescope, \cite{bol07} mapped a region covering the Bar and
part of the Wing of the SMC in the Spitzer Survey of the Small 
Magellanic Cloud (S$^3$MC).  The survey covered roughly 10 square 
degrees using all of the filters available on the Infrared Array 
Camera \citep[IRAC; 3.6, 4.5, 5.8, and 8.0~\mum;][]{faz04} and on the 
Multiband Imaging Photometry for Spitzer \citep[MIPS; 24, 70, and
160~\mum;][]{rie04}.  The IRAC observations were obtained in 2005
May. S$^3$MC was expanded on by the Surveying the Agents of Galactic Evolution program 
\citep[SAGE;][]{mei06}, which had initially observed the LMC.  
SAGE-SMC observed a 30 square degree area covering the entire
Bar and Wing of the SMC and extending east to the Bridge with all
seven IRAC and MIPS filters \citep{gor11}.  The temporal coverage
included two epochs, with the IRAC data in 2008 June and September.
After the end of the cryogenic portion of the Spitzer mission, the
SAGE-VAR survey obtained four additional epochs from 2010 August to
2011 June at 3.6 and 4.5~\mum\ in 3 square degrees centered on the 
Bar of the SMC \citep{rie15}.  

Our Spitzer program (Program ID number 13096) built on this legacy by adding two more epochs 
at 3.6 and 4.5~\mum\ covering the entire 30 square degree footprint 
of the SAGE-SMC survey in late 2017 and early 2018.  This program, 
entitled Spitzer's Last Look at the Small Magellanic Cloud, or
SMC-Last for short, when combined with the previous surveys,
provides a minimum of four epochs at 3.6 and 4.5~\mum\, creating a 
temporal baseline of over nine years covering the entire SMC and its 
surroundings.  Up to nine epochs are available in the core of the 
SMC spanning a period of over 12 years.

Section \ref{observations} describes the two new epochs of observations with Spitzer of
the SMC that form the basis of SMC-LAST.  Section \ref{overview} presents 
an overview of the resulting image products.  The following sections cover some of the
steps taken when processing the data, with corrections for background levels
in Section \ref{backgrounds}, corrections for artifacts in 
Section \ref{corrections}, masking for artifacts in
Section \ref{masking}, and other issues in Section \ref{misc}.  
Section \ref{discuss} provides some final thoughts.

\section{Observations} \label{observations}

We mapped $\sim$30 square degrees of the SMC, including the Bar, the Wing, 
and the Bridge that extends 
toward the LMC, at 3.6 and 4.5 \mum\ (Program ID 
13096) using Spitzer's Infrared Array Camera \citep[IRAC;][]{irac04}.
The Astronomical Observation Requests (AORs) were
based on those used for the 2008 maps for SAGE-SMC from \cite{sagesmc11}.
The observations were taken in the high-dynamic range mode (HDR), 
with 0.4 and 10.4 second integrations, which optimized the sensitivity to both 
bright and faint objects.

Each of the 29 primary AORs consisted of a 28$\times$14 raster with 
146\farcs4 and 292\farcs8\ steps (120 and 240 pixels) for the rows and columns,
respectively, covering regions that are $\sim$1\fdg1$\times$1\fdg1. 
The 29 raster scans needed for one iteration of the SMC map were intended to be
made with the same roll angle, and the two iterations of the map were to 
be executed three months apart. That would provide a 90\arcdeg\ roll, and 
the requested observation dates were chosen to avoid gaps within the maps.
This strategy gives a factor of
at least two in coverage in a single map, enabling removal of random
effects like cosmic rays and bad pixels, as well as systematics such as
latents from saturations or scattered light \cite[e.g.,][]{horaea04, horaea06,
horaea08irac}.
Because many of these artifacts preferentially affect either rows or columns
in the array, the 90\arcdeg\ roll from the second epoch allowed us to better
mitigate systematic effects such as column pull-down effects
from bright sources, as well as providing an additional factor of 2 in 
coverage.

\begin{figure} % Fig. 1
\includegraphics[width=0.9\textwidth]{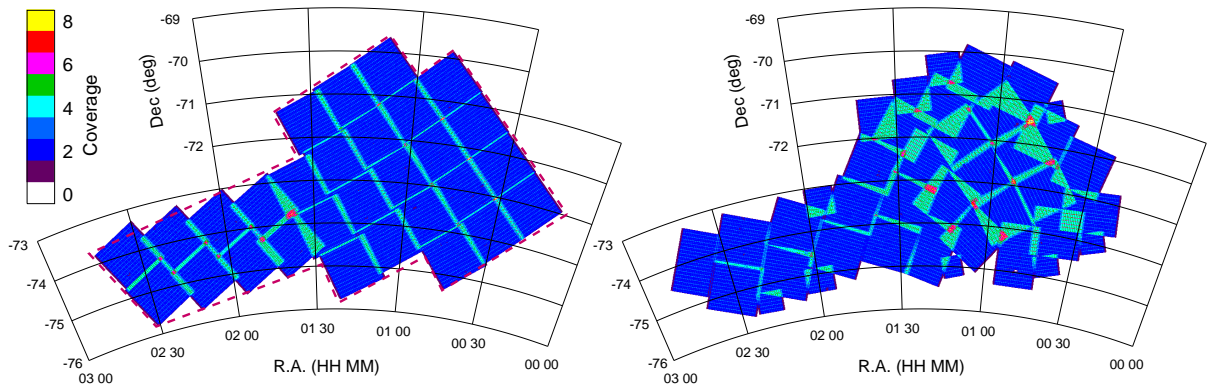}
\caption{
Final coverage of the SMC-Last program. (left) epoch 1, 2017 Aug 25 - Sep 13;
(right) epoch 2, 2017 Nov 24 - 2018 Feb 12.
The SAGE-SMC survey region is shown as the dark red dashed outline. ``Coverage'' is
the number of 10.4s integrations observing each location in the region.
\label{fig.irascov}
}
\end{figure}

For the first map (epoch 1), the observations were made largely as planned, taking
place over the course of 18 days, 2017 August 25 to 2017 September 13. A 
small rotation is present in the AORs in the Bridge relative to those nearer the core, 
from the rotation of the IRAC field of regard between execution of the 
initial and final AORs. Due to data-volume constraints and limited availability
of the ground station, the observations for the second epoch spanned 78 days, starting
2017 November 24 and ending 2018 February 12. The resulting change in roll
angle over this period left gaps among the planned AORs. Working with the Spitzer
Science Center enabled us to add several smaller AORs using Director's 
Discretionary Time to fill in many of those gaps. Figure
\ref{fig.irascov} illustrates the final IRAC coverage for the two epochs.

\section{Overview of image products} \label{overview}

\begin{figure} 
\includegraphics[width=0.9\textwidth]{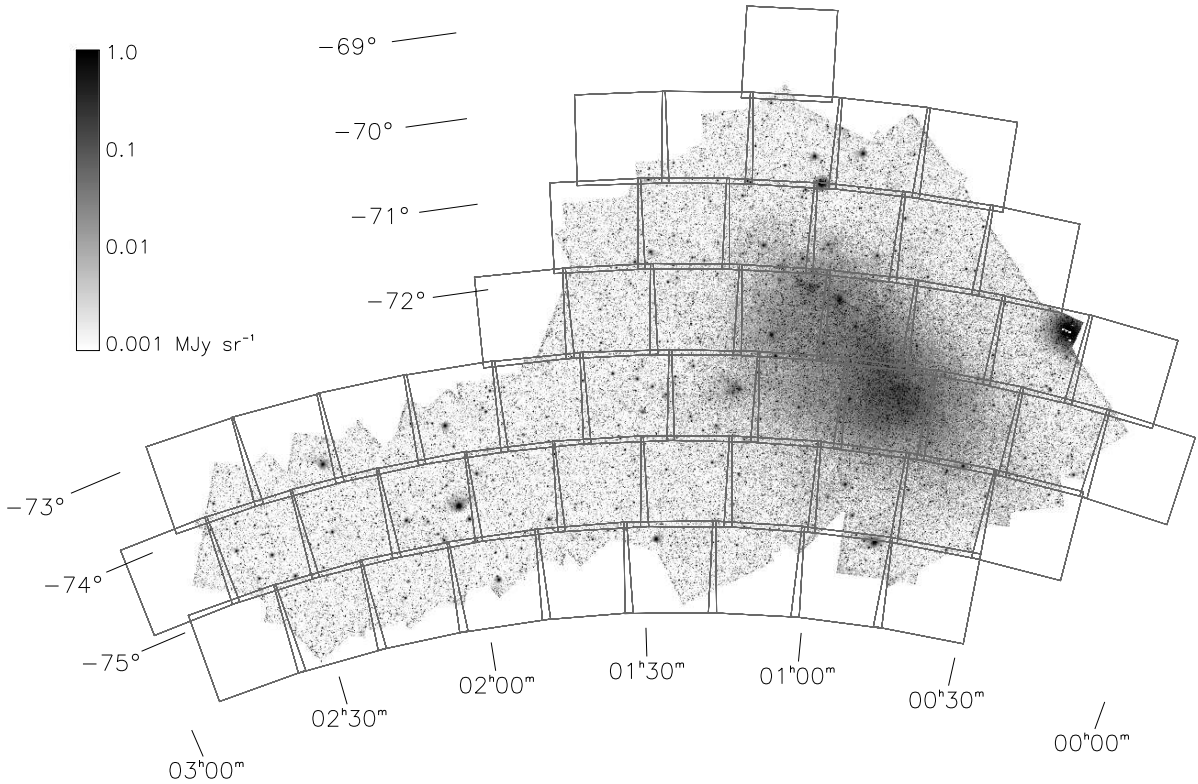}
\caption{
The 52 FITS image boundaries superposed on
a composite of the combined-epoch data for channel 2. \label{fig1}
}
\end{figure}

The SMC-Last mosaics comprise, for each epoch and channel, 52 FITS image files,
each covering a region approximately 1$\fdg$06 $\times$ 1$\fdg$06,
with an angular pixel size of 0${\farcs}$6. 
The images are made in tangent
projections in J2000 coordinates, with the projection center at the center
of each image, and a rotation angle of 0$\arcdeg$. The images are aligned
in rows 1$\arcdeg$ apart in declination (Decl.), and in each row the images are
approximately 1$\arcdeg$ apart in right ascension (R.A.) with the overall aligment of the
images chosen to best cover the survey region. The image-to-image overlaps
are 3$\farcm$5 in Decl. and a minimum of 1$\farcm$6 in R.A. Figure \ref{fig1} shows
a schematic of the image boundaries superposed on a composite image of
4.5 \mum\ data for the combined epochs.

The data comprise image sets for each of the two channels in both epochs, plus
each channel in a combination of both epochs. In addition to the image data,
each image file has corresponding FITS files containing coverage maps
(the number of data pixels influencing each mosaic pixel) and uncertainty maps.

Table \ref{summtable} provides a summary of the properties of the image products.

The resolution of the mosaics has been investigated by creating composite point 
source images for each plate, and fitting a 2D Gaussian function to the composite
images. The mosaic sources have been selected by mapping sources from the 
2MASS ``6x'' catalog \citep{cutri12} (and the 2MASS catalog \citep{skrut06} where 
the 6x catalog lacked coverage,
longward of about R.A. 01$^{h}$52$^{m}$), using sources in the K-magnitude range
12.0-17.0. The matching mosaic point sources are aligned with the centroid pixel
and averaged together. The 2D Gauss fitting yields median FWHM values of 
$\sim$2${\farcs}$0 over all the data, in both x and y, corresponding closely 
to the R.A. and Decl. axes, with a mosaic-to-mosaic rms of about 0${\farcs}$1.
The centroid pixel alignment does smooth the point source profile somewhat; we 
estimate a pixel-centered profile would have a FWHM $\sim$0${\farcs}$05 smaller.

The mosaics have been constructed using the mosaic utility in the Mopex
software package provided by the Spitzer Science Center \citep{mopex05}. 
The Corrected Basic Calibrated Data (CBCD) pipeline products provided
by the Spitzer Science Center have been used throughout as the input data source to 
construct the mosaics. The supplied CBCD data products
have been corrected for detector linearization and flatfielding, calibrated to
flux units (\mjysr), have had dark current
removed and the post-BCD pointing refinement applied (although see Appendix \ref{refine} for
details of the pointing refinement used for our data), and have been flagged for radiation hits,
saturation effects, latent images, stray light and other artifacts \citep{irachb21}
A zodiacal component has been removed with the dark frames, but not from the 
data, so the CBCD products have the difference included.
The zodiacal model values supplied with the
CBCD products indicate that the zodiacal levels typically vary by less than 0.005 \mjysr
over any given AOR, and this is absorbed into the background level adjustments described below.
The 10.4s integrations only have been used in
the mosaics, except where, as noted below (section \ref{satsub}), the
0.4s integration data are used to fill in saturated pixels for very bright
sources.

The options for coadding in the mosaics include weighting by the uncertainty 
values, which according to the Mopex handbook \citep{mopex05} is not recommended, 
weighting by integration time, or equal weighting. We have used equal weighting, 
but as we have used the 10.4s integrations almost exclusively, this is effectively weighting 
by integration time. The exception is the saturation substitutions, discussed in
section \ref{satsub}.

The CBCD products have been processed further prior to the construction
of the mosaics, as described in the remainder of this paper. 
In the following, the term ``pipeline'' refers to the CBCD processing pipeline, 
and ``BCD'' refers to a single frame of CBCD data for the 256$\times$256 detector arrays.
Channels 1 and 2 are the 3.6 and 4.5 \micron\ data respectively.

\section{Background level adjustments} \label{backgrounds}

\begin{figure} 
\includegraphics[width=0.9\textwidth]{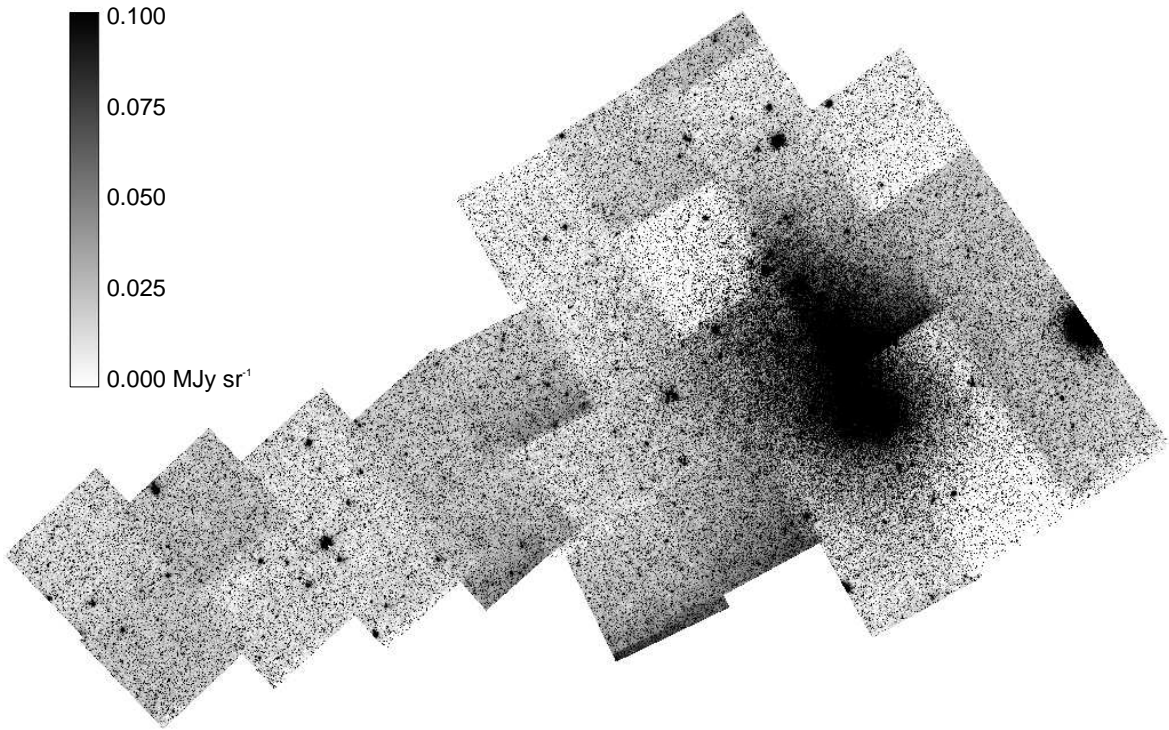}
\caption{
Composite of epoch 1, channel 1 raw CBCD data,
without the background level adjustments described in section \ref{backgrounds}. \label{fig2}
}
\end{figure}

Figure \ref{fig2}  shows a composite image of the epoch 1, channel 1 data using the
CBCD products without further processing. The scaling
is linear to emphasize the low-level backgrounds. It is apparent that there is a significant
non-astronomical variation in the background levels from AOR to AOR,
with an overall scatter of about 0.1 \mjysr. This variation indicates
that this data set does not yield absolute background levels or variations
at a length scale of the typical AOR coverage region (about a degree).
For the purpose of point source extraction from the mosaics we would like
to remove the discontinuities at the AOR boundaries.

If these level variations can be characterized by scalar offsets at the BCD
level (as opposed to a true gradient being added at either the BCD or AOR
level), then in principle the backgrounds can be brought to a common, self-consistent
value with an overlap-matching procedure, in which each BCD is given a scalar offset,
with the offset values simultaneously optimized with a least-squares algorithm
to minimize each BCD's overlap differences with its neighbors. In practice, however, an
overlap correction procedure that exactly minimizes the overlap differences
tends to produce large ``ramping'' excursions in background levels, because such a
routine responds to small systematic variations in BCD backgrounds, and there is
no constraint on the magnitude of the corrections.

Also apparent in Figure \ref{fig2} is that the AOR backgrounds not only have a generally scalar
differential offset, but typically also an overall
gradient, either increasing or decreasing in level over the duration of the AOR, plus in many cases
a steep increase or decrease in background levels at the start of the AOR.
The approach we take is to ``detrend'' the backgrounds for each AOR separately
(described in section \ref{detrend}), and then apply a modified version of the BCD overlap
matching procedure to resolve residual BCD-to-BCD level differences (section \ref{overlap}).
To mitigate potential ramping in the overlap-matching procedure, processing
is first applied to reduce systematic pixel response variations and artifacts that
persist for the duration of an AOR. This is described in the following section.

\subsection{Median-image subtraction}

\begin{figure} 
\includegraphics[width=0.9\textwidth]{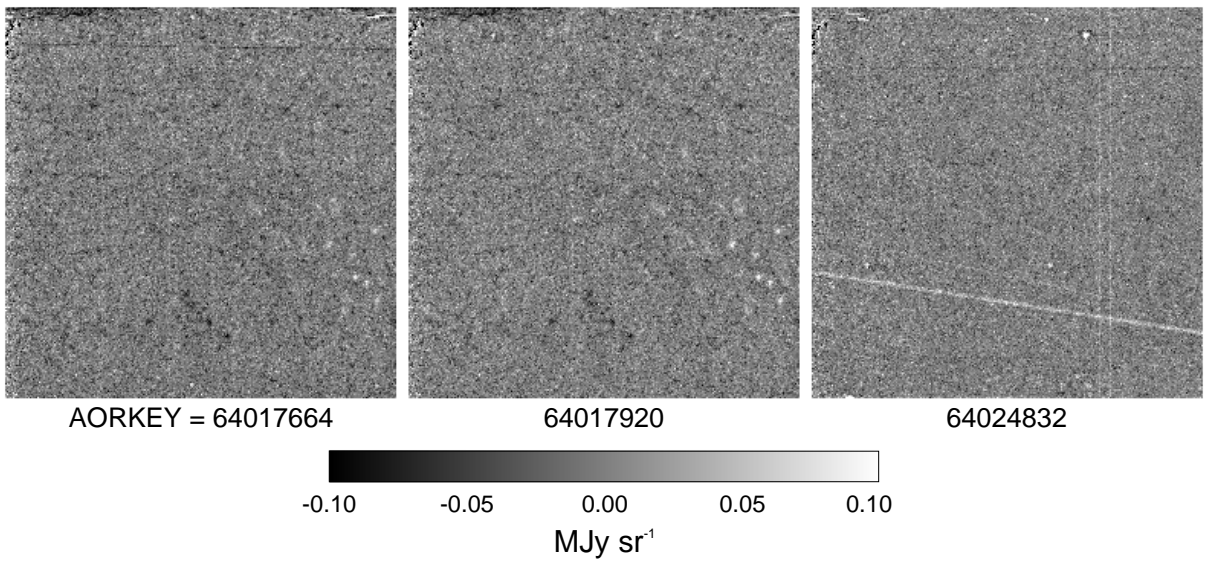}
\caption{
Examples of median images for three AORs, all
for channel 1. Typical are
the spots seen at lower right in the first two examples. These may be latent images of
bright point sources but they are not flagged by the CBCD pipeline. The streak in
the example on the right is likely the result of the array slewing across a very bright
source. \label{fig4}
}
\end{figure}

The median BCD array images for the AORs (i.e., the median value of each array
pixel over the duration of an AOR) typically show artifacts that are persistent for a
large fraction of the AOR. Figure \ref{fig4} shows a few examples. Also of
concern, for the overlap-matching procedure, are possible small systematic
variations across the array. To mitigate these effects, a form of the
median image is subtracted from each BCD: the AOR is divided into two
sections chronologically (i.e., for a 1-hour AOR, each section consists of
30 minutes of data). The median BCD image is calculated for each section,
and then the overall level of each is normalized by subtracting the median value of the image,
giving a mean value close to zero. A correction image is 
formed by assigning each pixel as the smaller magnitude of the two median images.
This correction image is visually compared to the first and last BCDs
in the AOR; if the artifact in the correction image represents a transient and is not
present in both BCDs, the process is repeated with the AOR divided into 3 or
4 sections as necessary to eliminate the transient artifact from the correction.

\begin{figure} 
\includegraphics[width=0.9\textwidth]{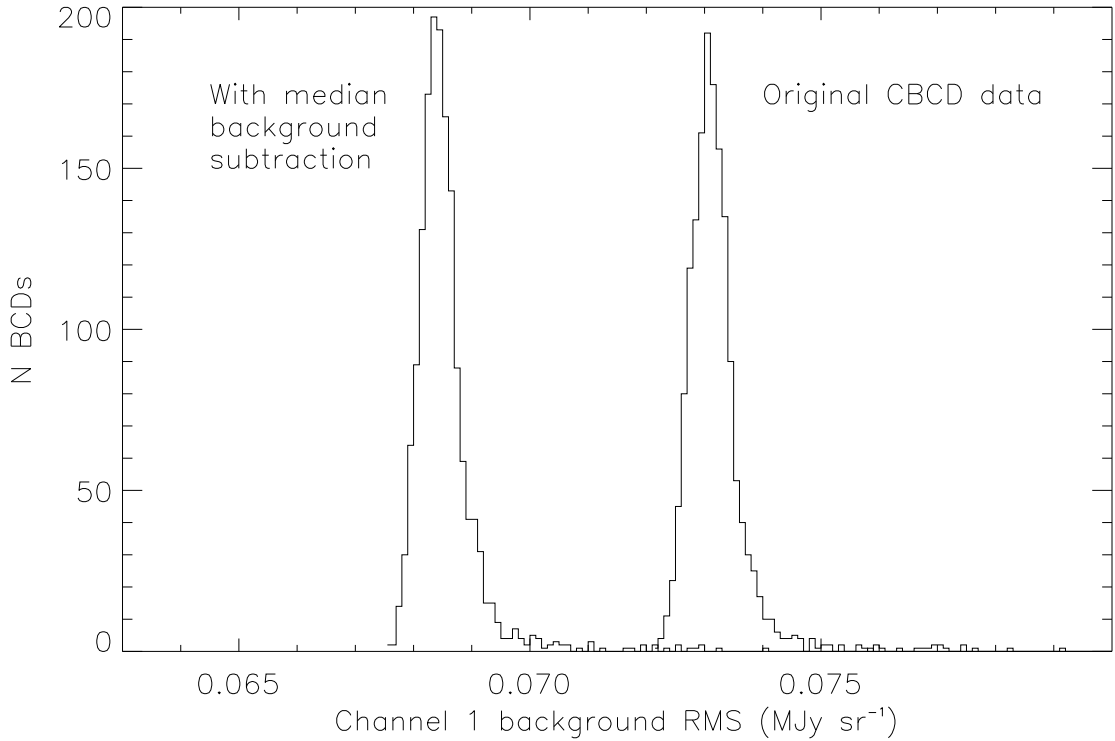}
\caption{
Histograms of background noise levels of $\sim$3000 BCDs from four AORs 
in the Bridge, for the original CBCD data, and the results of the median background 
subtraction. This is channel 1 data from the first epoch.\label{fignoise}
}
\end{figure}

Two concerns arise with this procedure. The first is that the correction images
have an apparent residual noise level, and so the subtraction could increase the uncertainty in
the BCD data. We have examined 4 AORs from each epoch in the Bridge region (where the
backgrounds are fairly flat) to investigate this question, comprising $\sim$3000 BCDs in
each band. For each pair of BCDs (original CBCD and median-image subtracted),
most sources, plus artifacts and bad pixels, are first deleted by eliminating pixels more than 
0.25 \mjysr\ from the median in either BCD, and then eliminating pixels within 
one pixel of those pixels. The rms of the remainder
is then calculated. For the 3.6 $\micron$ data, the original CBCD data have a median rms of 
0.073 \mjysr, and the median-image subtracted data have a median rms of 0.069 \mjysr.
For the 4.5 $\micron$ data, the results are 0.059 \mjysr\ and 0.055 \mjysr\ respectively.
Figure \ref{fignoise} shows a histogram of the results for the 3.6 $\micron$ data in 
the first epoch.  We find thus that the rms noise levels in the BCDs
actually decrease slightly, by about 5-7$\%$.

\begin{figure} 
\includegraphics[width=0.9\textwidth]{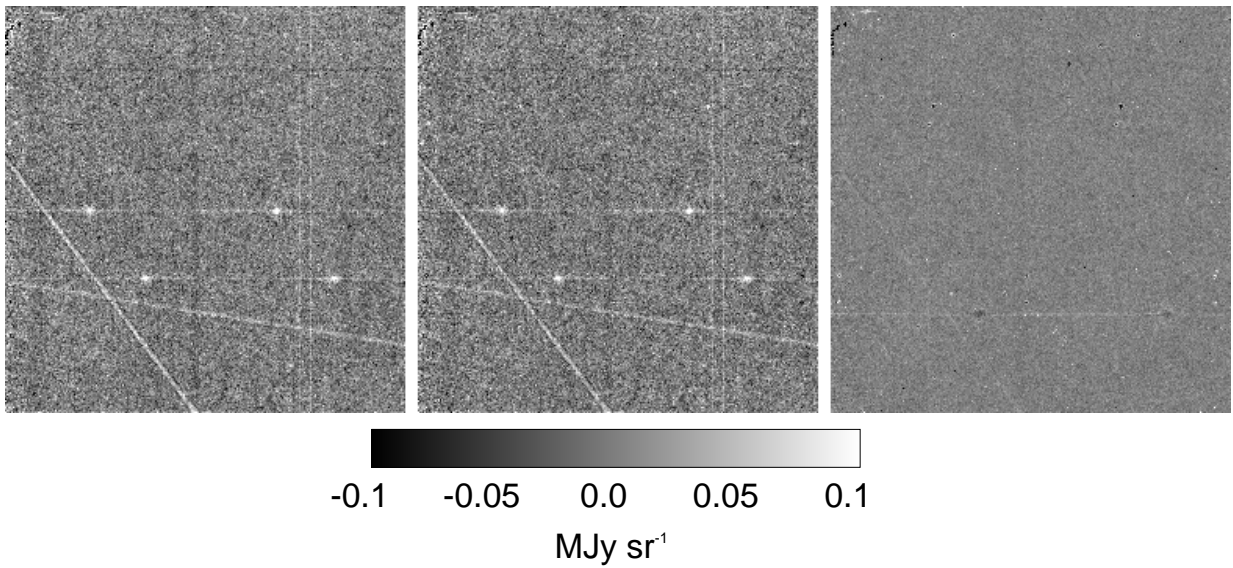}
\caption{
Median images created from the BCDs over first half (196 BCDs) of AORKEY=64024064, 
channel 1 (left panel), the
second half (middle panel) and the difference between the two (right panel).  
The rms for the median images is about 0.033 \mjysr\ and for the difference
image about .014 \mjysr. If the apparent noise in the median images were true
random noise, we would expect the rms of the difference image to be about 0.047 \mjysr.
\label{fignoise3}
}
\end{figure}

To investigate whether this decrease could be due to deletion of artifacts
rather than an actual reduction in the background noise, Figure \ref{fignoise3} shows
median images constructed from the first half of the BCDs from AORKEY=64024064 (channel 1)
and from the second half. The third panel shows the difference between the
two. If the rms reduction were due to removal of artifacts, and the background
``noise'' in the median images were true noise, the difference 
image would show a noise level increased by about $\sqrt{2}$. To the contrary,
the rms of the difference image is reduced by more than a factor of two 
($\sim$0.033 \mjysr\ in the median images and $\sim$0.014 \mjysr\ in the difference
image.) 
This indicates that the ``noise'' in
the median correction images is primarily some true systematic pixel-to-pixel variation 
consistent throughout the AOR.

The second concern is that this correction will tend to remove true sky
gradients from each BCD, if such a gradient is uniformly present across
the region covered by the AOR. While subtraction of an overall sky gradient
will not occur in this step (because each BCD in an AOR is given the same
normalized background correction), BCD-level flattening could potentially produce
a ``staircase'' background artifact, although we see no evidence of this in the
mosaics.

\subsection{Detrending} \label{detrend}

In this procedure, the scalar BCD background levels for each AOR are 
adjusted separately to remove
overall trending in the levels. The median of each BCD in the AOR is taken, and the
results are initially fitted with a function consisting of an exponential plus a linear term:
\begin{equation}
y(x) = ae^{\frac{-x}{b}} + cx + d
\end{equation}
where the ordinate values $x$ are the scaled sequence number of the BCD in the
AOR, and the coefficients are determined
with a nonlinear least-squares algorithm, using uniform weighting for all the
data points. Then, for values of $x$ greater than 3 times the ``time constant'' $b$ 
(to exclude the exponential portion),
the data values above the fitted curve are given zero weight, and the values below
are weighted by the distance below the curve, with the weights normalized to the
maximum distance. The curve is then refit with the adjusted weights, and then the
weights are adjusted again similarly with the refit curve, giving both first and 
second refits.

The intent
is to create an estimate of the background bias function for the AOR, where the
assumption is made that the observations for any given AOR will spend a significant fraction of the
time covering portions of the sky at some ambient background (away from the SMC
core) which we are arbitrarily assigning zero brightness, noting as above that this
data set is not expected to preserve low level large-scale extended emission.

\begin{figure} 
\includegraphics[width=0.9\textwidth]{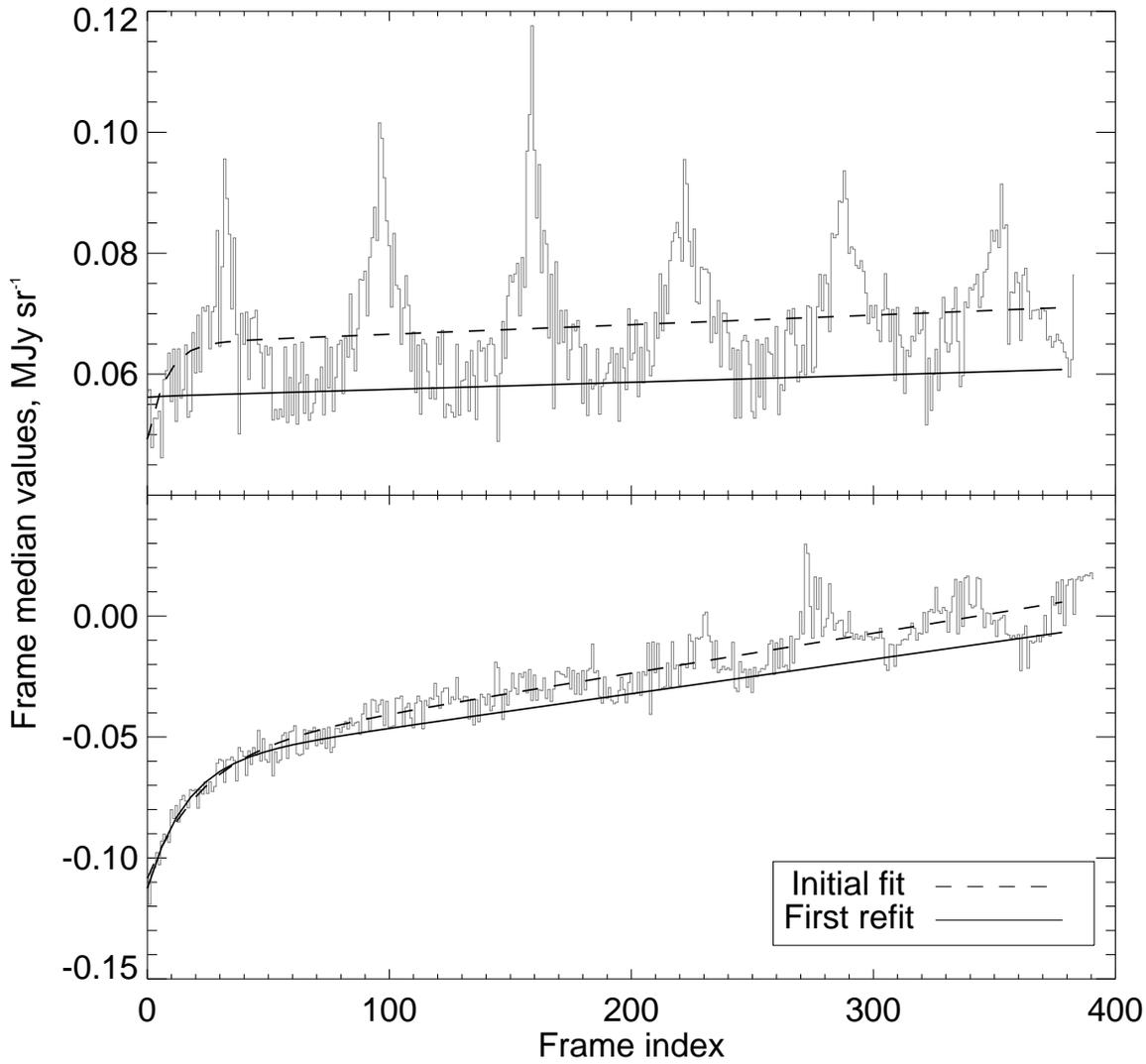}
\caption{
Two examples of the BCD-median functions, the
initial fit, and first refit as described in the text. Both are channel 1. The upper case is an 
AOR (AORKEY=64021760) that periodically crosses
the SMC core, and the lower (AORKEY=65213184) is in a region with flatter backgrounds. \label{fig5}
}
\end{figure}

The intended function is thus one that skims the lower bound of the BCD-median
data. The first refit usually provides the appropriate results, and in a few
cases the second refit is used. Also, in a few cases where the fitting failed,
the linear portion of the curve is assigned by hand. Figure \ref{fig5} shows two examples
of the BCD-median function, the initial fit, and the first refit, for an AOR that crosses the
SMC core, and one in a region with flatter backgrounds. The sinusoidal appearance of the
median data for the first case is caused by the raster
scanning repeatedly over the core.

For each BCD in the AOR, a scalar background bias value is calculated from the fitted
curve and subtracted.

The issue for this procedure is to what extent it is affecting relative
extended emission and background levels, particularly at the $\sim$1\arcdeg\ length scale of
the AOR-sized regions. (Much smaller extended features should be unaffected
by the simple linear background fitting, and we do not expect to recover
background features at much larger length scales.) This will be addressed
in the next section.

\subsection{Overlap level matching} \label{overlap}

Following the detrending, all the BCDs (for a given channel) are processed with
an overlap matching procedure \citep{mizuno08}, in which each BCD is given a (typically small)
scalar offset to least-squares minimize the residual level differences between
overlapping BCDs, to reduce or effectively eliminate background discontinuities
in the mosaics. This procedure is described by \cite{mizunoea08b}.
This is a ``damped'' overlap algorithm intended to suppress the ramping effect
common with an exact overlap matching algorithm. In essence, the magnitude
of the correction for a given BCD is included in the calculation of $\chi^{2}$
in the least-squares algorithm, along with the overlap differences with its neighbors.
The cost of the ramping suppression is, in principle, small residual level
differences in the BCD-to-BCD overlap regions, but this has not been observed
in practice.

Both epochs for each channel are overlap-matched simultaneously, to
give common background levels between the epochs, noting that true background
levels (apart from zodiacal differences, which we are not attempting to preserve) 
are not going to change observably over the timescale of the two epochs.

\begin{figure} 
\includegraphics[width=0.9\textwidth]{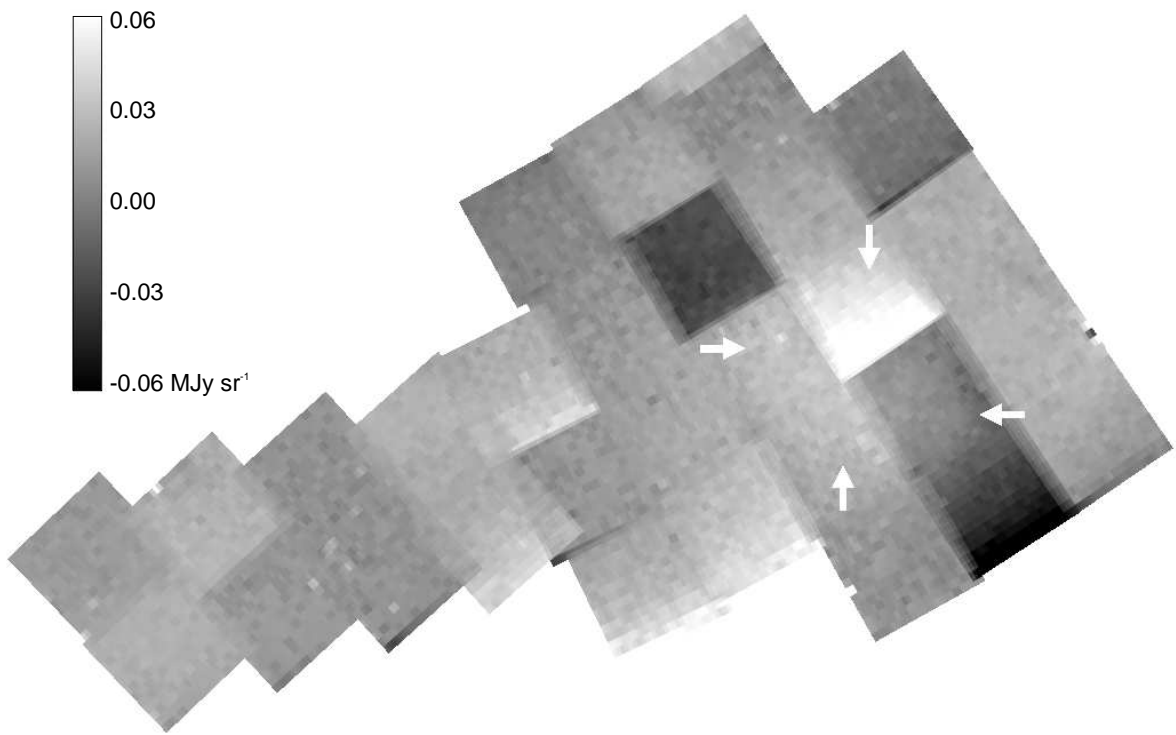}
\caption{
Difference image between the raw CBCD data
(i.e. Figure \ref{fig2}) for epoch 1, channel 1, and the data processed with the detrending
and overlap matching procedures. The arrows show the apparent boundary of the SMC core. \label{fig3}
}
\end{figure}

Figure \ref{fig3} shows the difference 
between a composite image of the unprocessed data for epoch 1, channel 1 (i.e.,
Figure \ref{fig2}) and the data processed with the detrending and overlap
adjustments described above. (The median-image subtraction is omitted because
it does not affect overall background levels, but the systematic pixel value
adjustments over each AOR cause visible features along the scan rasters in the
difference image that obscure subtle background level differences).
Ideally we would see nothing but level variations that align with
the AOR boundaries, and this is generally the case. The exception is that
the backgrounds in the core of the SMC appear to be slightly oversubtracted;
the arrows in the figure indicate the faint boundary visible. The magnitude
of the apparent oversubtraction is around 0.01-0.02 \mjysr. This effect
is absent for the case where the detrending alone is included, so it seems
to be a consequence of the overlap matching procedure. While the detrending
procedure would be expected to depress backgrounds for AORs that have
higher true mean backgrounds than their neighbors, the overlap matching
procedure is not expected to have any particular overall bias, and so we
cannot necessarily conclude that these results are a relative
oversubtraction of the background levels in the SMC core. The
alternative is to postulate that there is some systematic elevated bias in the raw
data for the SMC core that roughly scales with the true background
levels. We note that the IRAC channel 3 and 4 Si:As detectors had just such
a scattered-light issue, and could also be present at low levels for the InSb
detectors of channels 1 and 2. 

\section{Artifact corrections} \label{corrections}

\subsection{Column pull-down and pull-up corrections} \label{pull}

The column pull-down artifact is a phenomenon in which entire columns that contain a very bright
source are depressed in intensity. 
The nature of the depression varies from very uniform
along the affected columns, which allows for a correction, to highly irregular, which can
only be masked, and even very uniform examples typically show some irregularity at the
very top and bottom of the array. The width of these artifacts generally scales with the 
brightness of the source, and somewhat surprisingly, the wider artifacts from very bright 
sources tend to be more uniform than the narrower examples.

While the CBCD pipeline includes a correction for column pulldown, examples 
remain either uncorrected or incompletely corrected, and we address these residual
artifacts.

Column pull-up is a similar phenomenon in which
entire columns have an elevated intensity. In contrast to the pull-down artifact,
however, pull-up artifacts have extremely uniform elevated levels along the columns and
can nearly always be corrected. For these cases, there is seldom
any obvious triggering source, although many seem to be associated with a point source at the
very top or bottom of the array, and others seem to be associated with the first latent images of bright 
point sources.
The column pull-up artifacts can be from one to a few columns in width
and in some cases are up to about a quarter of the array wide (these band-like
artifacts are almost always associated with first latent images).

The correction for these artifacts exploits the circumstance that the backgrounds in most of
the survey region are nearly flat, generally lacking any significant structured extended emission.
The basic approach is to apply a scalar offset to the regions of depressed or
elevated columns to match the overall median for the BCD. Specifically, the procedure is
to identify the boundary columns of the depressed or elevated regions, thus dividing the
BCD into groups of columns with uniform level offsets. The median of each group is given
a scalar offset to match the global median for the BCD, with boundary columns
adjusted individually to match the global median. Note that while
narrow artifacts are corrected to the presumably non-elevated background levels over
the rest of the BCD frame, for the
bandlike pull-up artifacts, no assumption is made about the ``truth'' background, and
the correction just sets the column region level offsets to a common value. The subsequent
background overlap matching procedure described above adjusts the background levels to agree with
the BCD's neighbors.

\begin{figure} 
\includegraphics[width=0.9\textwidth]{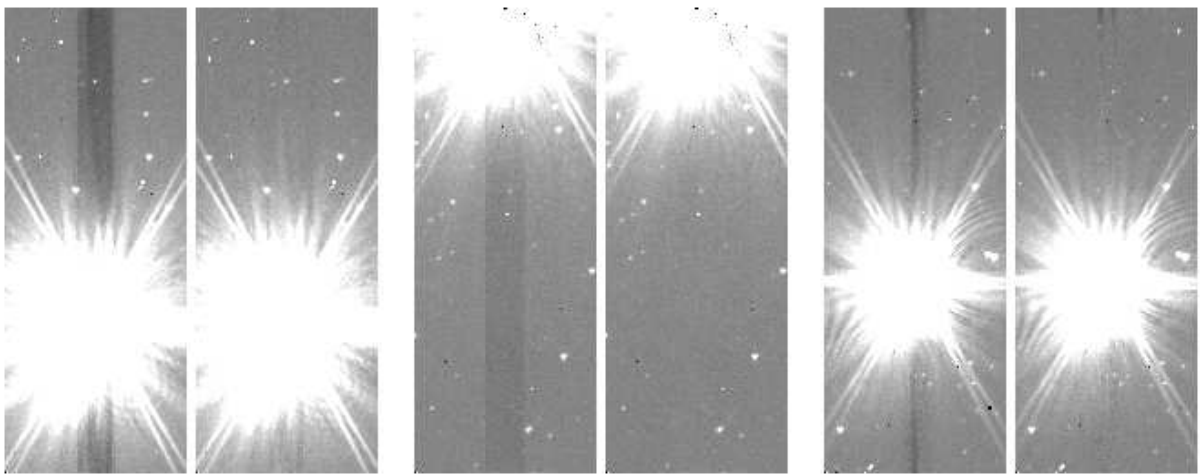}
\caption{
Examples of the pull-down artifact and their
corrections. Note that the correction for the case on the right leaves a residual at the top
of the array; this is subsequently masked. \label{fig7}
}
\end{figure}

\begin{figure} 
\includegraphics[width=0.9\textwidth]{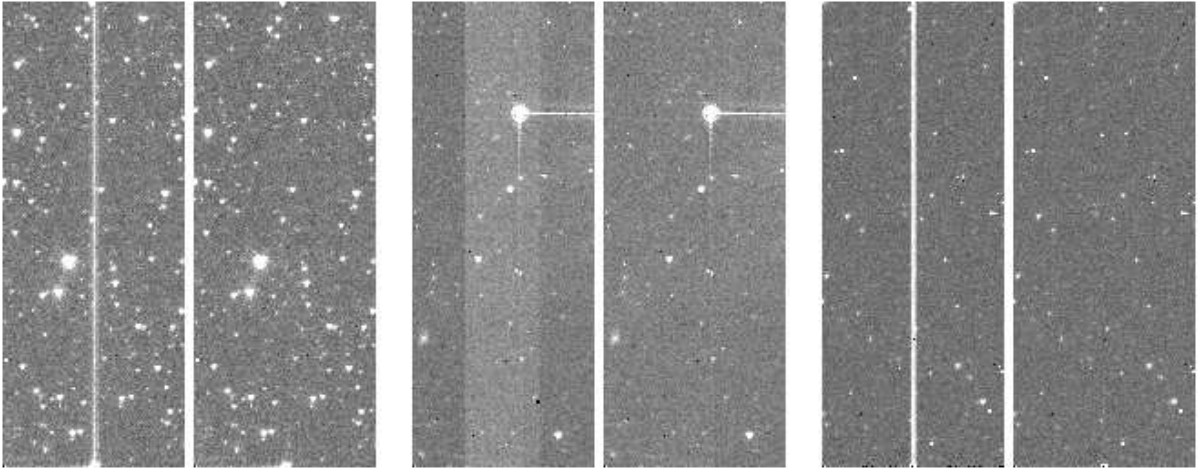}
\caption{
Examples of the pull-up artifact and their
corrections. The bright source in the band-like artifact at center is the first latent
of a bright point source. Note that the left and right cases are associated with faint point
sources at the bottom of the array. \label{fig8}
}
\end{figure}

Very narrow pull-up artifacts are easily identified and corrected in software, if the
elevation is about 0.05 \mjysr\ or more. 
The remaining pull-up and all pull-down artifacts are corrected, if possible, by hand as 
they are encountered from inspections of the mosaics and BCDs. The by-hand corrections also allow
ad-hoc masking of irregular portions of the pull-down artifacts.
Figure \ref{fig7} shows a few examples the pull-down artifacts and the resulting corrections,
and Figure \ref{fig8} likewise for pull-up artifacts.

\subsection{Latent corrections}

In channel 1, latent images from bright point sources can persist for minutes or tens of
minutes (i.e., dozens or hundreds of subsequent BCDs). For channel 2, nearly all latent
images fade to undetectability within approximately a minute (in this data set). The CBCD
pipeline masks the latent images, but in channel 1, for the long duration latents, 
the CBCD pipeline also commonly continues the latent masking long after the latent has 
become undetectable. 

Also, the pipeline misses many instances of latents, in both channels,
particularly when they fall near the very edge of the array. The cutoff appears to be
about 10 pixels from the edge. For the 3.6 $\micron$ data, 
over all the AORs in the survey, the pipeline has flagged 634 latent sequences further than 10 pixels
from the array edge; we would therefor expect about 112 sequences ($\sim$15\% of the 
total) nearer the edge, whereas 
the pipeline has flagged 20. For the 4.5 $\micron$ data, similarly, the pipeline flagged 
1190 sequences further than 10 pixels from the edge, and only 24 nearer the edge, where we
would expect more than 200 (and the pipeline flagged none within 10 pixels of the
top and right edges).

The long-duration channel 1 latents typically decrease in brightness rapidly and nonlinearly
for the first few BCDs and then fade nearly linearly until they become undetectable. This linearity
is exploited to apply a correction: for a given latent image occurrence in the linear portion of
the sequence, the latent itself is estimated as the median value of the affected region of
the array over a 9 BCD window centered on the given BCD, with an appropriate scalar background 
subtraction. The latent image estimate is then subtracted from the BCD array, and the pipeline masking
for that latent image is deleted.

The latent correction procedure is applied with an interactive routine: for each latent sequence,
regions for the correction and background estimation are selected (the pipeline tends to
underestimate the size of the latents for very bright sources), then the corrections are
applied and the results inspected {\em in situ} on each affected BCD, to assess the 
quality of the correction. Generally the correction is effective starting at the fifth
latent occurrence.

The visual inspection process also permits two additional latent correction steps:
cases in which the pipeline masking has been extended long after the latent has faded, and
for these the masking is simply turned off; and cases in which the pipeline misses
identifying latents. For these, the short-duration cases are simply masked, and
the correction procedure is applied to the long-duration cases. 

While channel 2 does not have long-duration latents, the channel 2 BCDs are also inspected
to identify and mask the pipeline-missed latents. 

The visual inspections have found about half (64) of the ``expected'' edge latents for channel 1,
and about a quarter (49) for channel 2 (plus, for both cases, a roughly equal number 
scattered over the array). 

\begin{figure} 
\includegraphics[width=0.9\textwidth]{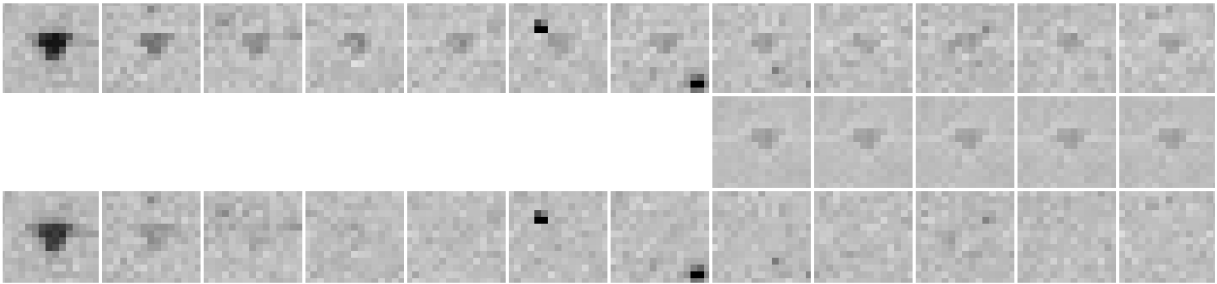}
\caption{
Example of a sequence of channel 1 latent images and
the corrections. Top row: the initial latent and the 11 subsequent images at that location
in the array, following observation of a $\sim$3 Jy point source. In this case, the latent
images persist for about 20 minutes. Middle row: the
calculated latent images as described in the text. The first calculated latent is at
the eighth occurrence in the sequence, and is provisionally applied to the earlier occurrences. Bottom row:
the affected regions after the subtraction of the latents. The first applied correction is
at the fifth occurrence; the earlier latent images remain masked. \label{fig11}
}
\end{figure}

Figure \ref{fig11} shows an example of a sequence of latents, the calculated latent images,
and the subtracted results.

\subsection{Large-scale latents}

When an extremely bright point source crosses the arrays (during slewing between frames),
it can leave a streak-like latent image in the array that can persist for as long as the
longest point-source latents. These cases are not flagged by the CBCD pipeline. For these
latents, a procedure similar to the point-source latent corrections is applied: the
boundary of the latent itself is defined, and for each latent image, the 9 BCD window
median image is taken for the whole array, and, again exploiting the generally flat
backgounds in this data set, the median of this image, apart from the latent region itself, is
regarded as the scalar background and subtracted. The resulting latent image is subtracted
from the BCD, and similarly visually inspected for the effectiveness of the correction.

\section{Artifact masking} \label{masking}

\subsection{Stray light}

The stray light artifacts are patches of light, usually seen in a few distinctive
patterns, that are confined to about the upper third of the array, presumably
due to bright sources just off the upper edge of the array. The CBCD pipeline makes
an attempt to predict and mask these artifacts, but it identifies only the
brightest cases, and the selected masking region is typically much larger than
necessary. The pipeline-flagged artifacts have been visually examined and the
masking regions reset manually as necessary to accommodate the actual size and
extent of the artifact.

\begin{figure} 
\includegraphics[width=0.9\textwidth]{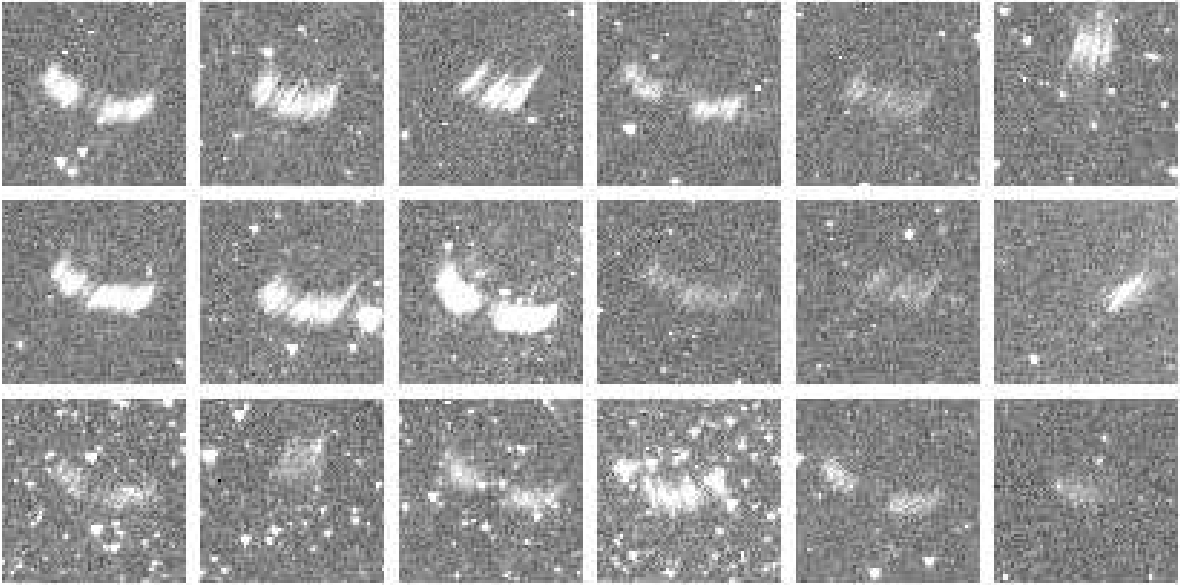}
\caption{
Some examples of the stray light artifact.
The top row are cases caught by the CBCD pipeline, and the remainder were missed. Note
that the pipeline misses some very bright cases. \label{fig6a}
}
\end{figure}

There has otherwise been no systematic effort to identify the remaining (mostly
fainter) stray-light artifacts. However, the distinctive shapes are readily apparent
in the mosaics and individual BCDs, and are thus masked by hand when they are
located. Figure \ref{fig6a} shows a few examples of stray light, both caught
and missed by the CBCD pipeline.

The IRAC Instrument Handbook indicates that there is also a small region off the lower
edge of the channel 1 array that can cause stray light artifacts. No cases
of this have been noticed, although faint examples could have gone undetected.

\subsection{Column pull-down}

While a portion of the column pull-down artifacts are correctable using the
procedure described in section \ref{pull}, for the majority the residuals after the
correction are sufficiently irregular to warrant simply masking the artifact
instead. Towards that end, all point sources that are either saturated or above
a peak threshold (300 \mjysr\ for channel 1 and 200 \mjysr\ for channel 2) are
visually inspected on the BCDs for the pull-down artifact, and if present,
evaluated for whether correction or masking is indicated, and then respectively applied.
If masked, the masking boundaries are selected by hand, but the triggering source itself
is generally left unmasked. 

\begin{figure} 
\includegraphics[width=0.9\textwidth]{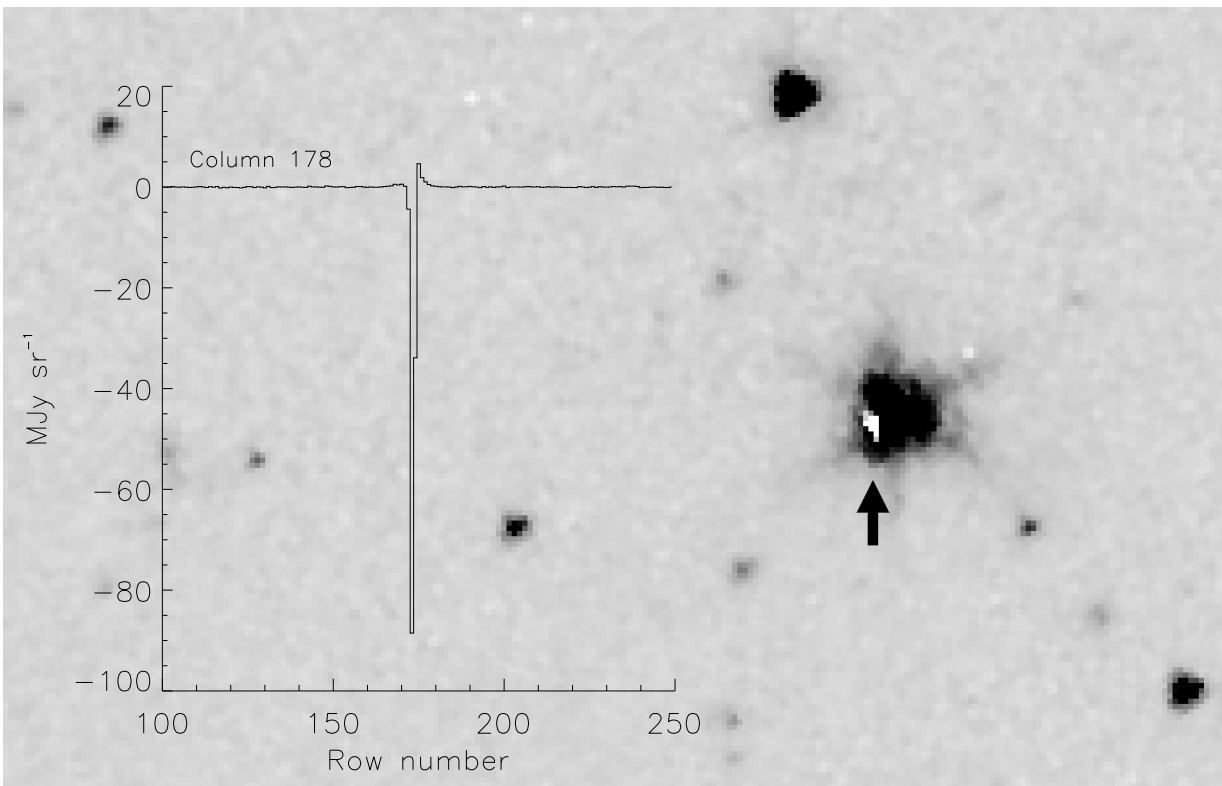}
\caption{
Example of the very-negative pixel values
seen in some ``pull-down'' columns, and their effect on resulting mosaic images. Note that
this is not a data hole in the mosaic, rather a very negative value, because the outlier
rejection mechanism fails to trap these cases. The affected source is at
$\sim$02$^{h}$00$^{m}$, -74$\arcdeg$52$\arcmin$. \label{fig10}
}
\end{figure}

A second artifact is found in cases in which, in a fraction of the columns labeled
as pull-down by the CBCD pipeline, one to a few contiguous pixels have values
far below the local backgound, from one to tens of \mjysr. These usually
occur in columns that have been flagged as pull-down and sufficiently corrected in the 
pipeline, as they show no residual pulldown effect. 
These artifacts are usually not caught in the
outlier rejection mechanism in the mosaic construction, and so leave
small negative ``holes'' in the mosaics. These artifacts are identified as
pixels beyond a specified threshold level (0.5 \mjysr) below the local background level
for ``pull-down'' labeled columns, and masked. Figure \ref{fig10} shows
an example of a column with this artifact and the uncorrected effect on the resulting
mosaic image.

The cause of this artifact appears to be a consequence of the pipeline pulldown 
correction itself, as the corresponding uncorrected BCD does not show this
artifact.

\subsection{Charged-particle strikes}

Particle strikes can create an artifact in which a portion of the
column in which the strike occurs, up to about 50 pixels above and below
the strike, are corrupted such that one side is elevated and the other
depressed. While the strike itself is usually either flagged and masked
in the pipeline or in the outlier rejection in the mosaic creation, the
corrupted pixels are not, and can leave an artifact in the mosaics. These
cases are masked by hand as they are identified.

\begin{figure} 
\includegraphics[width=0.9\textwidth]{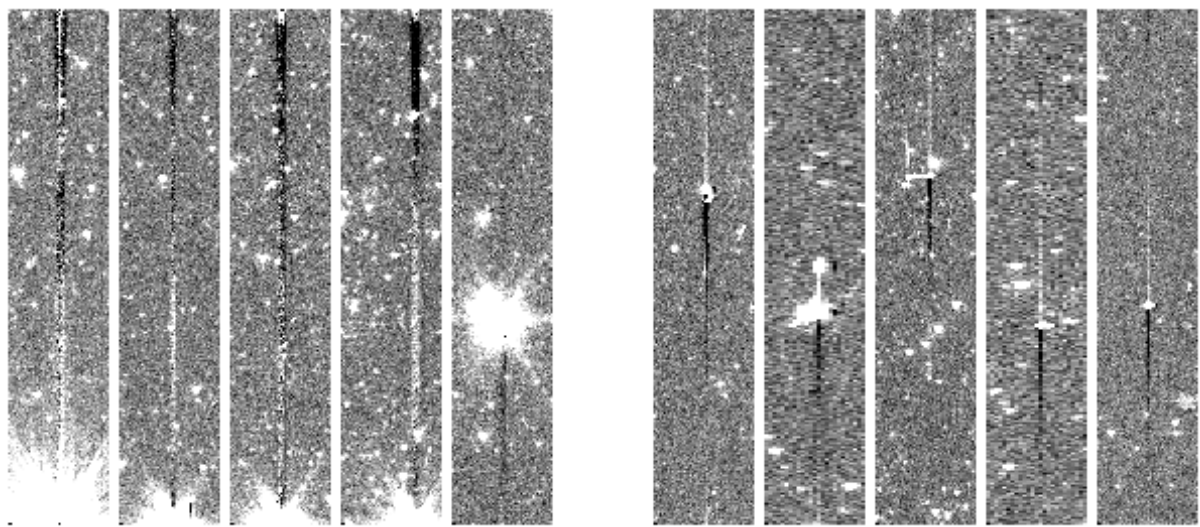}
\caption{
Examples of uncorrectable pull-down
artifacts (left) and charged-particle strikes (right). \label{fig9a}
}
\end{figure}

Figure \ref{fig9a} shows a few examples of both uncorrectable pull-down artifacts
and these charged-particle strikes.

\section{Miscellaneous corrections and residual artifacts} \label{misc}

\subsection{Saturation substitution} \label{satsub}

The BCD pixels flagged and masked by the pipeline as bright point-source saturations
have been replaced with the corresponding pixels from the short-integration (0.4s)
data taken contemporaneously with the long-integration data. For this procedure,
the short-integration pixel values are inserted into the long-integration BCD arrays
prior to the mosaic construction, along with substitution of the corresponding pixel
values for the uncertainty data. (The pointing from the 0.4s to the 10.4s
integrations seems to be sufficiently stable for the pixel-for-pixel substitution.)
Note that we are effectively using integration time as the weighting in the mosaic
construction, and so this pixel substitution will result in improper weighting for
the substituted 0.4s pixels. However, regions saturated in a given 10.4s BCD are unlikely 
to have unsaturated 10.4s data in any overlapping BCDs, and so the improper weighting
is unlikely to result in improperly combined data in the mosaics.

The saturations due to true point sources are distinguished from saturations from
other causes (typically charged-particle strikes on the array) by examining the data
in a 4-pixel boundary around the saturation: for saturated point sources the
median value of the boundary region is always 2.0 \mjysr\ or higher above
the local background whereas for charged-particle strikes it is always well
under 1.0 \mjysr.

\subsection{Pixel mask augmentation}

The bad pixels in the channel 1 and channel 2 arrays are supplied as pixel masks in the set of
standard CBCD pipeline data products for each AOR. For the time periods covered by the SMC-Last
observations, these pixel masks are constant for both channels. However, it has been found
through inspection of the mosaics and the contributing BCDs that, for channel 2, a number
of pixels behave erratically at least over the course of an AOR, but are not flagged in the
nominal pixel mask. While anomalous high values for a pixel are usually removed in the
outlier rejection procedure, anomalous low values typically are not and result in ``holes''
in the mosaics.

These bad pixels are identified and added to the nominal pixel mask. For each AOR, the
median value of each array pixel is calculated, giving the median array image for that
AOR. Pixels that differ from the median of that image by at least 0.15 \mjysr\ are
taken to be bad pixels (the rms of the AOR-median images is typically less than
0.01 \mjysr). This gives an ad-hoc bad pixel mask for that AOR. In the generation
of the mosaics, for each mosaic a new bad pixel mask is created that is the union of
the pixel masks for each AOR included in the mosaic and the nominal pipeline pixel mask.
The nominal channel 2 pixel mask has 97 bad pixels flagged; the augmented pixel masks
typically have 10-50 additional pixels flagged.

\subsection{Anomalous outlier rejection} \label{reject}

In the mosaic construction, we have used the outlier, dual outlier, and box outlier options
available in the Mopex utility. We have found that these outlier rejection mechanisms 
together generally work well for the typical coverages in our data.
However, this outlier rejection has an anomalous behavior
in which, if a mosaic pixel is covered by data for two BCDs, and there are additionally two or more
BCDs covering the mosaic pixel which have been masked at that location, the rejection mechanism
tends to regard all local maxima above a very small amplitude, and the immediately
surrounding regions, as outliers and deletes them from the mosaics. The cause of this anomaly is unclear,
but we have applied a mitigation procedure separately for the
single-epoch and combined-epoch mosaics.

For the single epoch mosaics, sets of mosaics were created for both channels and both epochs with
the outlier rejection mechanism turned off entirely. The standard mosaics were then compared
to these specially prepared mosaics. Where the standard mosaic showed a data hole, as indicated
by a not-a-number (NaN) value in the coverage image, and the prepared mosaic showed data
present, this was considered a candidate for transferring the data from the prepared mosaic
to the standard mosaic. An interactive routine, with a visual inspection of the regions from
the two mosaics, was used to distinguish true cases of the anomalous rejection from actual
outliers. For the true cases, the data (for image, coverage, and uncertainty mosaics) were
transferred, for a region including the NaN ``hole''
and also a two-pixel boundary surrounding the hole, because a BCD pixel anomalously rejected
covering the hole can cause a data deficit for mosaic pixels up to two pixels away from the hole,
given that the mosaic pixel size is half the angular size of the BCD pixels.

For the combined-epoch mosaics, the mosaics were similarly compared with these
rejection-reprocessed single epoch mosaics. Where a data hole in the combined-epoch
mosaic corresponded to data present in the single-epoch mosaics, the data were transferred,
also with a two-pixel boundary around the hole. If data were present in one of the single-epoch
mosaics, but not both, the appropriate data were simply transferred to the combined-epoch
mosaic. If data were present in both epochs, the image pixel values were calculated as the
weighted average of the corresponding single-epoch pixel values, with the weights being the
coverage values in the single-epoch mosaics (noting as in section \ref{overview} that we 
are effectively using
integration time as the weighting in the mosaics rather than the uncertainty values).
The coverage is simply the sum of the individual coverages. The resulting uncertainty
for the combined-epoch mosaic pixels is also a weighted combination. For uncertainties
$\sigma_{1}$ and $\sigma_{2}$ for the two epochs, and coverages $c_{1}$ and $c_{2}$, the combined
uncertainty is:

\begin{equation}
\sigma_{combined} = \frac {\sqrt{ c_{1}^{2}\sigma_{1}^{2} + c_{2}^{2}\sigma_{2}^{2} } }{c_{1} + c_{2}}
\end{equation}

\subsection{Background striations}

\begin{figure} 
\includegraphics[width=0.9\textwidth]{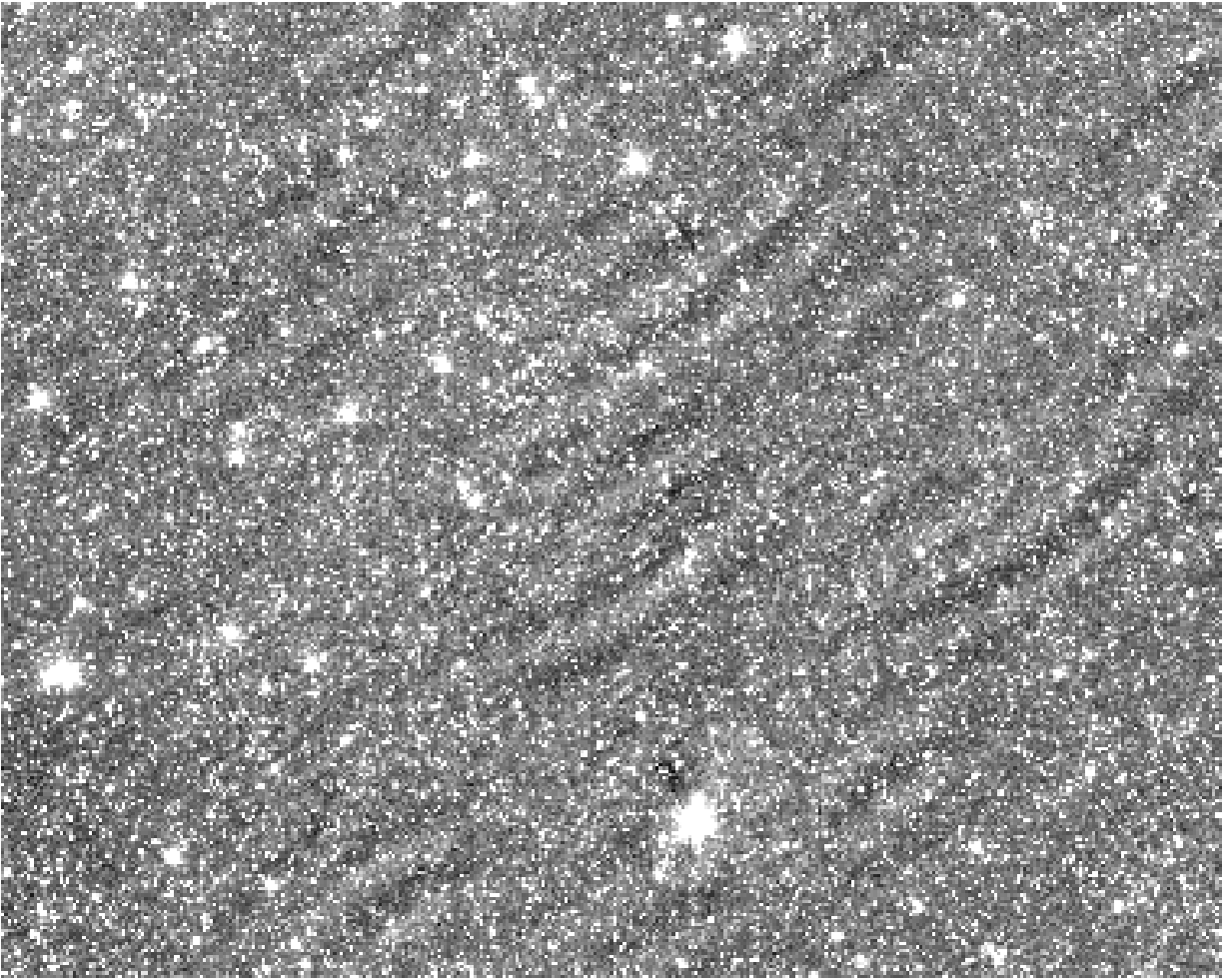}
\caption{
Example of striation pattern artifact in the
mosaics. This is from the epoch 1, channel 2 data. The field is 1$\arcdeg$ wide, and
centered at 00$^{h}$52$^{m}$, -70$\arcdeg$54$\arcmin$. The amplitude of the
striations is about 0.01-0.02 \mjysr. The ``culprit'' is AORKEY 64022016.  \label{fig12}
}
\end{figure}

Several of the mosaics show a distinctive striation pattern in the backgrounds. Figure \ref{fig12}
shows an example for epoch 1, channel 2. Where they occur, they occur to some extent in both
channels, and occur in both epochs, albeit in different locations. The amplitude is around
0.01-0.02 \mjysr. The striations are not sufficiently regular to allow a correction,
so they have been left unaddressed.

It appears that the striations are confined to a limited set of AORs, seven in epoch 1 and
four in epoch 2, and originate
from irregularities in the backgrounds in the BCDs, roughly horizontal patterns that persist for a
few successive frames and then evolve to a different pattern. As the horizontal patterns are also
parallel to the scan direction, this gives rise to the extended features in the mosaics.

Table \ref{stritable1} shows the affected AORs with the date and start UT.

\section{Discussion and conclusions} \label{discuss}

The SMC-Last program surveyed the SMC at 3.6 and 4.5 \mum\ in two epochs, the
first from 2017 August to September, and the second from 2017 November to 2018 February.
We have created sets of 52 1\fdg1 $\times$ 1\fdg1 images of the SMC from IRAC 
3.6 and 4.5 \mum\ observations. The data were corrected for instrumental artifacts
and background discontinuities prior to mosaicking. The processed data result
in six mosaics, three in each filter, with one from each of
the two epochs and a third from the combined epochs. The final mosaics are
available as FITS files from the Infrared Science Archive (IRSA).

The combined data sets from S$^3$MC, SAGE-SMC, SAGE-VAR, and SMC-LAST
provide four epochs with the 3.6 and 4.5~\mum\ IRAC filters covering the
entire SMC and a fifth in the Bar, and four more in the center, which
totals between four and nine epochs covering a temporal baseline
of nine years everywhere in the SMC and 12 years in the center.  This
temporal coverage opens a rich variety of research avenues ranging from
astrometric studies in the foreground to variability studies in the SMC
and the background.

\begin{figure}
\includegraphics[width=0.9\textwidth]{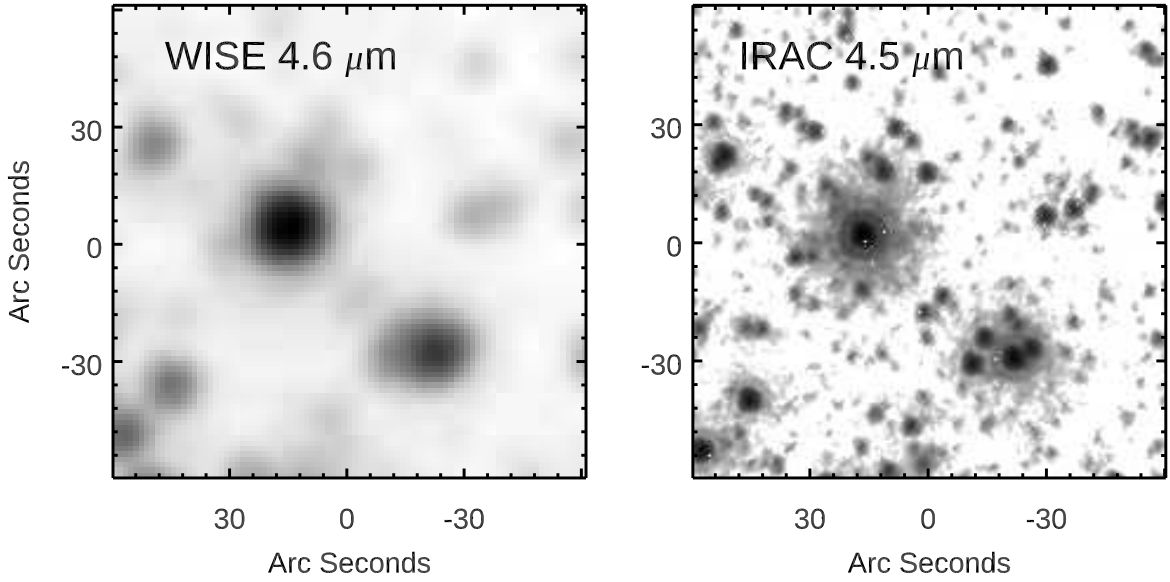}
\caption{Comparison of WISE and IRAC, a 2\arcmin\ box centered on
($\alpha,~\delta$)=(12.4121, $-$73.0397). (left) WISE 4.6 \mum; (right) IRAC
4.5 \mum.
 \label{figwise}
}
\end{figure}

Data from the Wide-field Infrared Survey Explorer \cite[WISE;][]{wri10},
currently operating as the NEOWISE-R mission \cite{mai14}, provide even
more temporal coverage.  WISE observes the SMC every six months in two
filters with similar wavelengths to the 3.6 and 4.5~\mum\ IRAC bands.  These
data can be combined with the IRAC data, although some differences must be
considered.  The WISE filters differ slightly from the IRAC filters and
are centered at 3.4 and 4.6 \mum.  Their angular resolution is
$\sim$6\arcsec, compared to $\sim$2\arcsec\ for the IRAC data during 
the post-cryogenic mission \cite[][v.\ 4, their Table 2.1]{irachb21}.  The limiting
magnitudes at 3.4 and 4.6~\mum\ for the cryogenic WISE mission were 16.5 and 15.5
mag \citep[respectively,][]{wri10}, and $\sim$15.5 and 14.3 mag for the
post-cryogenic mission \cite{mai14} for single-epoch photometry.  The
SAGE-SMC project, performed during Spitzer's cryogenic phase, had
corresponding limiting magnitudes of 17.6 and 17.0 mag \cite{gor11}. 
The SMC-Last point source extraction is still in progress (Kuchar et al. in
preparation), but the preliminary estimates for the limiting magnitudes
are $\sim$17.1 and 16.8 mag.  Figure \ref{figwise} compares a small
region of the SMC in the two 4 \mum\ bands.

The WISE observations come at a steady cadence of one epoch every 180
days (with a gap from 2010 to the beginning of the NEOWISE-R mission in
2014).  With the release of data from 2021, the extended WISE mission provides a
total of 20 epochs.  For targets which can be observed with WISE, the
temporal coverage in the SMC now extends to between 13 and 16 years,
with additional epochs expected.  One issue to be considered is the steady
six-month cadence, which leads to potential aliasing problems.  For
example, studies of long-period variables can be hampered because one year is
close to the pulsation period of many Mira variables \citep{whi94}.  The IRAC
observations can break that degeneracy.

The SMC-Last data will build on the legacy of the previous studies of
the infrared variability of long-period variables (LPVs) in the SMC
\citep[e.g.,][]{rie15}.  Multi-epoch optical studies have identified
$\sim$20,000 LPVs across the entire galaxy \citep[e.g.,][]{sos11}, but
as stars die, they embed themselves in optically thick dust shells and
become too faint for detection in the optical.  These heavily enshrouded
and optically invisible stars dominate the observed dust production in
the SMC \citep{boy12,sri16}, but the optical multi-epochs studies miss
them.  Similar objects in the Large Magellanic Cloud are barely at the
edge of detection with WISE \citep{slo16, gro18}.  To study them in the
SMC, the greater sensitivity of SMC-Last and other Spitzer-based surveys
is required.

The additional epochs from SMC-Last will also facilitate the study of
young stellar objects (YSOs) in the SMC.  The variability of T Tauri stars has been
recognized for decades \citep{joy45}, and more recent studies reveal
that nearly all YSOs vary at some level \citep[e.g.,][]{meg12,ric12,ric15}.
The lower metallicity of the SMC leads to more luminous YSOs compared to
the Galactic counterparts, and the lower dust abundances make them
visible earlier in the evolution \citep{wit03}.  Most of the
star-forming regions in the SMC were included in the S$^3$MC survey, so SMC-Last
provides a 12-year baseline to study this variability and search for
transient events like FU Ori-like eruptions.  SMC-Last will also enable
searches for transients in the background population of galaxies.

Astrometry of the foreground population can reveal brown dwarfs due to
their large proper motions.  Assuming a limit of 17.0 mag at 4.5~\mum,
SMC-Last can detect early L dwarfs to 275 pc, early T dwarfs to 115 pc,
and early Y dwarfs out to 25 pc \citep{kir11,kir12}.  We estimate,
based on space densities from \cite{cru03} and \cite{kir12}, that a
comparison of SMC-Last to SAGE-SMC should reveal the motion of
many dozens of L dwarfs, $\sim$10 T dwarfs, and 0.3 Y dwarfs in the
field of 30 square degrees.

Thus the two new epochs of infrared photometry from SMC-Last create many
new temporal research opportunities in the SMC, behind the galaxy, and
in the foreground.

\appendix

\section{Pointing refinements} \label{refine}

\begin{figure} 
\includegraphics[width=0.9\textwidth]{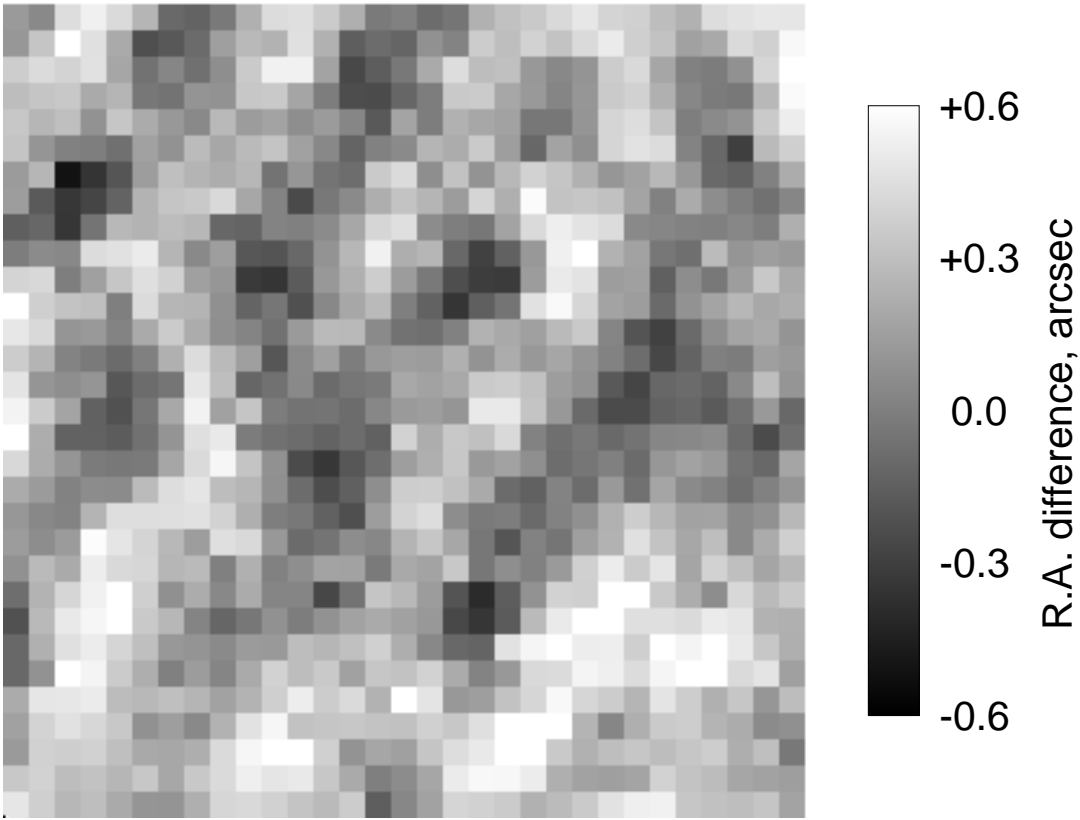}
\caption{
R.A. differences between epoch 1 and epoch 2 sources, with the ``superboresight''
pointing, for the $\sim$1$\fdg$1$\times$1$\fdg$1 mosaic centered 
at 01$^{h}$31$^{m}$, -73$\arcdeg$22$\arcmin$. The
sources have been binned in 2$\arcmin$\ bins and the image values calculated as the 
median of the R.A. differences in the bin. This is channel 1 data, but channel 2
shows a nearly identical pattern.
\label{pfig1}
}
\end{figure}

The mosaics were initially constructed using the ``superboresight'' pointing
solution, i.e. the nominal astrometry information in the BCD headers, which the
IRAC instrument handbook indicates should have an rms accuracy of about 0$\farcs$16. 
However, in analyzing source extractions from the initial mosaics, it was
found that the measured source position differences between the two epochs, particularly 
the R.A., typically vary in curiously systematic patterns. Figure \ref{pfig1} shows
the R.A. differences between epoch 1 and epoch 2 channel 1 sources for the mosaic centered
at 01$^{h}$31$^{m}$, -73$\arcdeg$22$\arcmin$. 
For this image, the R.A. differences are binned in 2$\farcm$0 bins and the 
bin values are the median difference values in the bin. Most mosaic regions show a 
generally similar pattern in the R.A., to varying degrees in
both epochs, but not for Decl. with the exception noted below. The patterns for
channel 1 and 2 are nearly identical. 

\begin{figure} 
\includegraphics[width=0.9\textwidth]{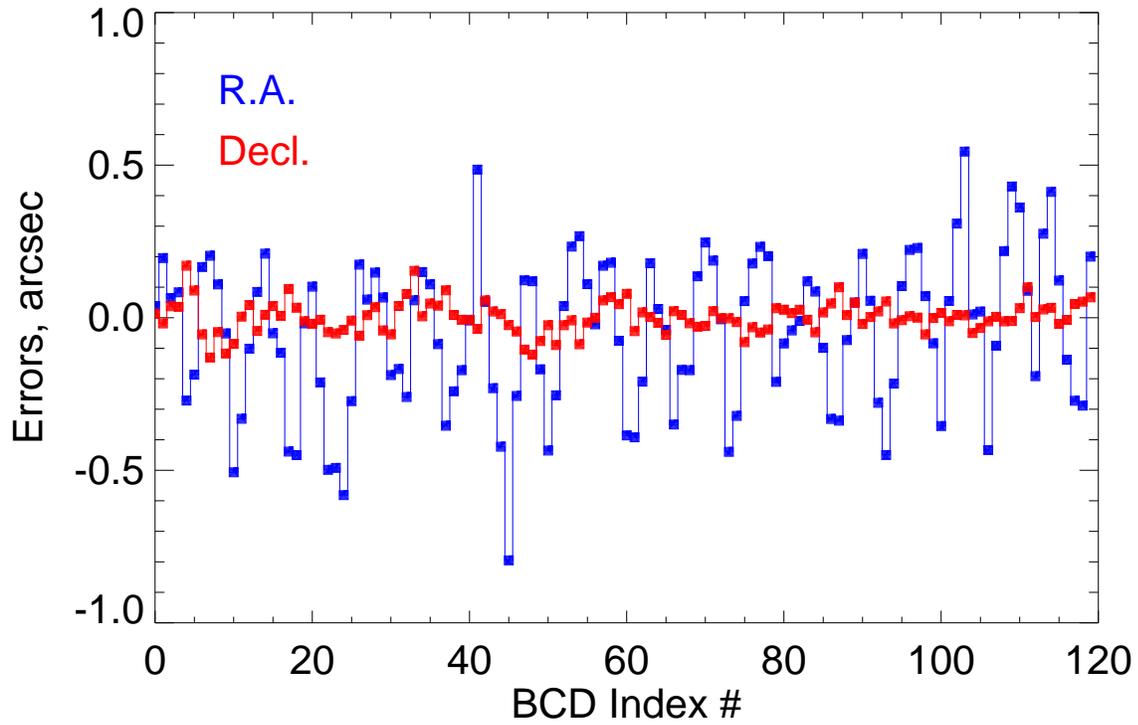}
\caption{
R.A. and Decl. errors for the first 120 BCDs of AORKEY=64019968, channel 1,
using the ``superboresight'' pointing solution,
showing the quasi-periodic error in R.A.
The errors have been calculated by matching 2MASS 6x sources to sources on
each BCD, and taking the median difference over the matched sources. 
The spacing between BCDs is about 26s.
\label{pfig2}
}
\end{figure}

The cause of these patterns has been found to be quasi-periodic errors in the
center R.A. (i.e. CRVAL1 keyword) in the BCD headers. Figure \ref{pfig2} shows
the errors for the first 150 BCDs for AORKEY 64019968, channel 1. The absolute 
errors have been calculated by mapping 2MASS 6x sources \citep{cutri12} into the astrometry for
each BCD, matching to the sources in the BCD image, and taking the median of 
the differences in R.A. and Decl. over all the matched sources. As noted above, 
the R.A. errors for channel 2 are nearly identical.
 
All of our AORs but six include the ``superboresight'' pointing. The six exceptions
are epoch 2 AORs covering the SMC core. For these AORs, both the R.A. and Decl. 
errors show similar periodicity and amplitude, and generally have larger 
global pointing offset errors as well, up to about 0$\farcs$5.

\begin{figure} 
\includegraphics[width=0.9\textwidth]{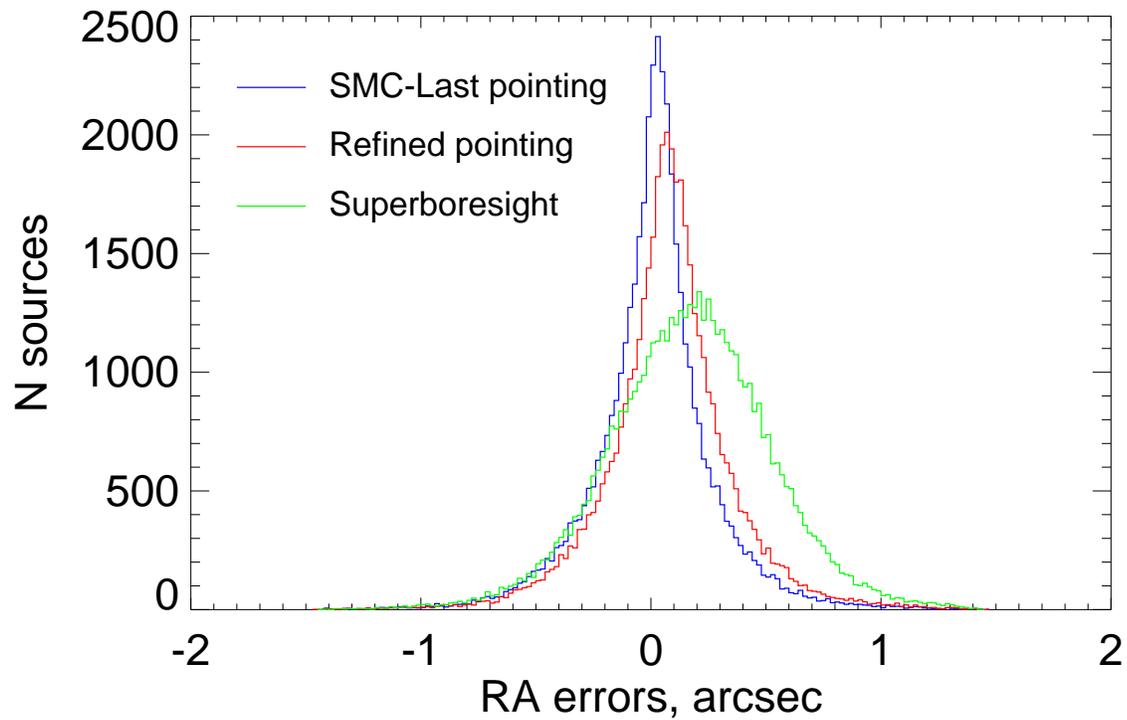}
\caption{
Histogram of R.A. errors for the SMC-Last pointing solution and the ``refined''
solution, for the mosaic centered at 01$^{h}$07$^{m}$, -72$\arcdeg$22$\arcmin$,
for epoch 2, channel 1.
The errors are calculated by matching the sources extracted from the mosaic
with the 2MASS 6x sources. The ``superboresight'' data included for comparison are from 
epoch 1 data for this mosaic, because the epoch 2 data for this mosaic did not include that 
pointing solution.
\label{pfig3}
}
\end{figure}

Most of our CBCD data, as provided by the Spitzer Science Center, include the
``refined'' pointing solution, for which the astrometry of
each BCD has been corrected to absolute coordinates by matching the sources on
the BCD frame to sources from the 2MASS catalog, and resetting the parameters
(CRVAL1, CRVAL2, and the CD matrix) to minimize the differences. For our data,
the ``refined'' pointing solution shows a significant improvement over the superboresight 
pointing. However, a small
quantity of our data is lacking the refined solution, so we have developed a
procedure to correct the absolute coordinates for each BCD using the 2MASS 6x catalog, 
and applied it to all our data. The details of the procedure are presented in 
Kuchar et al. (in preparation).  Figure \ref{pfig3} shows results of source 
extractions from mosaics created using our pointing correction procedure, and 
from the ``refined'' 
solution, with errors determined from matching to the 2MASS 6x catalog sources. 
Also shown in the figure 
for comparison are results for the ``superboresight'' pointing.

\begin{figure} 
\includegraphics[width=0.9\textwidth]{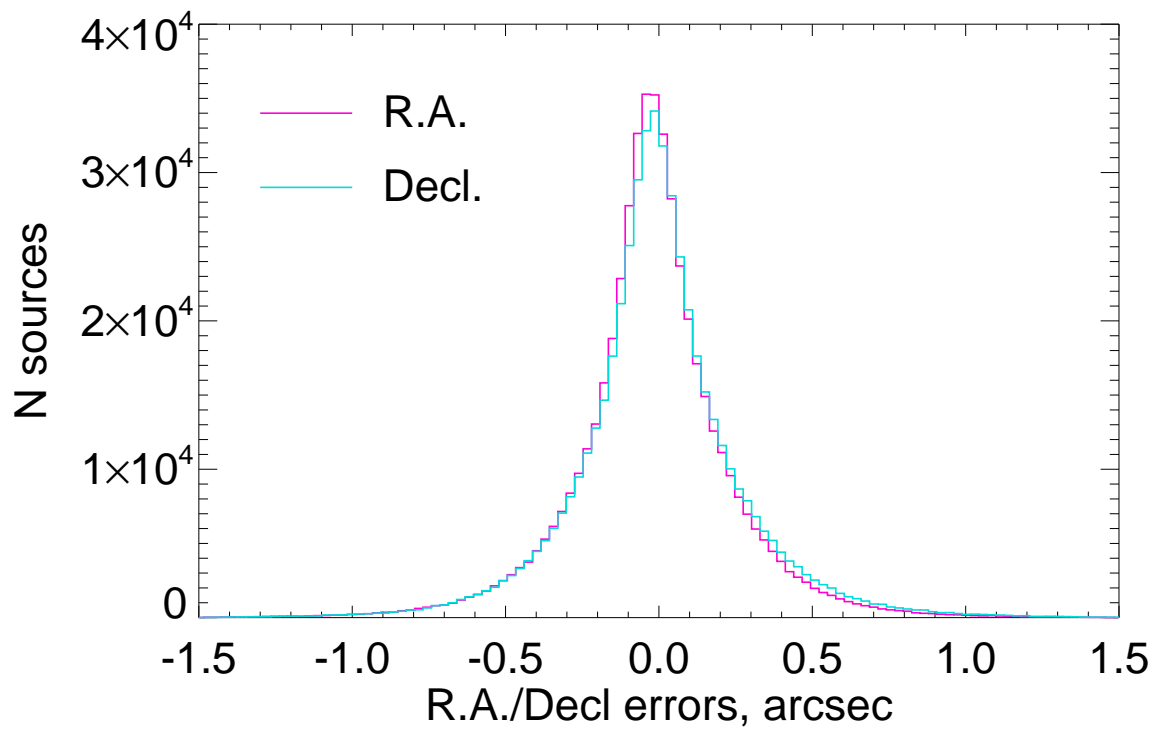}
\caption{
Histogram of R.A. and Decl. errors for the SMC-Last pointing solution for 
all epoch 1, channel 1 sources matched to 2MASS 6x (about 510,000 sources). 
The other epoch and channel results are similar. The median errors are all 
under 0$\farcs$03, and the rms values are $\sim$0$\farcs$25.
\label{pfig4}
}
\end{figure}

Figure \ref{pfig4} shows both R.A. and Decl. errors for the results of our pointing solution
for all of the epoch 1, channel 1 mosaics.

\acknowledgments{
This work is based on observations made with the {Spitzer Space
Telescope}, which is operated by the Jet Propulsion Laboratory,
California Institute of Technology under NASA contract 1407. We particularly
thank Nancy Silbermann of the Spitzer Science Center for her 
assistance in scheduling the ``fill-in''  observations.
Financial support for this work was provided by NASA through NASA ADAP
grant 80NSSC19K0585.}

\begin{deluxetable*}{lccc}
\tablecaption{SMC-Last Mosaic Properties\label{tab.props}}
\tablewidth{0pt}
\tablehead{
\colhead{Property}  &\colhead{Epoch 1} &\colhead{Epoch 2} & \colhead{Combined}
}
\startdata
Dates & 2017 Aug--Sep & 2017 Nov -- 2018 Feb & \nodata \\
Median Coverage Depth & 2 & 2 & 4 \\
Pixel-pixel noise\tablenotemark{a} \hspace{1pt} I1: & 0.043--0.10 & 0.044--0.092 & 0.032--0.091  \\
\hspace{7pt} (MJy sr$^{-1}$) \hspace{21pt} I2: & 0.034--0.050 & 0.033--0.058 & 0.024--0.054\\
Background level\tablenotemark{a} I1: & $-$0.005 -- +0.006 &$-$0.008 -- +0.005 & $-$0.007 -- +0.006 \\
\hspace{7pt} (MJy sr$^{-1}$) \hspace{21pt} I2: & $-$0.003 -- +0.002 &$-$0.001 -- +0.005 & $-$0.002 -- +0.004 \\
\cutinhead{Global Properties}
Pixel size  & \multicolumn{3}{l}{0\farcs6} \\
Angular resolution & \multicolumn{3}{l}{$\sim$2\arcsec} \\
Observed Area  & \multicolumn{3}{l}{30 deg$^2$} \\
No. of Plates & \multicolumn{3}{l}{52} \\
Plate size  & \multicolumn{3}{l}{1\fdg06 $\times$ 1\fdg06} \\
Bands & \multicolumn{3}{l}{3.6 \mum\ \& 4.5 \mum} \\
Integration Mode & \multicolumn{3}{l}{High-Dynamic Range} \\
Integration Times\tablenotemark{b} & \multicolumn{3}{l}{t$_{exp}$=0.6 \& 12 sec, t$_{int}$=0.4 \& 10.4 sec}
\enddata
\tablenotetext{a}{Pixel-pixel noise: standard deviation in a 51x51 pixel box
(30\arcsec$^2$) in a source-free region. Background level: mean
level in the same box. Estimated from one region in the SMC Bridge
and one just outside the Core.}
\tablenotetext{b}{In HDR mode, the two nominal exposure times of 0.6 and 12 sec lead to integration times of 0.4 and 10.4 sec.}
\label{summtable}
\end{deluxetable*}

\begin{deluxetable*}{lcccc}
\label{stritable1}
\tablecaption{AORs with background striations\label{tab.stri}}
\tablewidth{0pt}
\tablehead{
\colhead{AORKEY}  &  &\colhead{Date} & & \colhead{Start UT (hh:mm)}
}
\startdata
64018176 & 2017 & Aug & 25 & 00:19 \\
64017920 & 2017 & Aug & 25 & 03:20 \\
64020480 & 2017 & Aug & 29 & 02:10 \\
64020224 & 2017 & Aug & 29 & 05:01 \\
64019968 & 2017 & Aug & 29 & 07:52 \\
64021760 & 2017 & Sep & 02 & 11:06 \\
64022016 & 2017 & Sep & 05 & 08:52 \\
65213952 & 2018 & Jan & 07 & 15:37 \\
65215488 & 2018 & Jan & 14 & 12:42 \\
65215744 & 2018 & Jan & 21 & 13:11 \\
65255936 & 2018 & Feb & 08 & 17:03
\enddata
\label{stritable}
\end{deluxetable*}

% figcov1a.eps  fig01

% fig1.eps  fig02

% fig2.eps  fig03

% fig4.eps  fig04

% fignoise.eps  fig05

% fignoise3.eps  fig06

% fig5.eps  fig07

% fig3.eps  fig08

% fig7.eps  fig09

% fig8.eps  fig10

% fig11.eps  fig11

% fig6a.eps  fig12

% fig10.eps  fig13

% fig9a.eps  fig14

% fig12.eps  fig15

% i2w2comp_inv_2amin.eps fig16

% figpoint4.eps  fig17

% figpoint3.eps  fig18

% figpoint1.eps  fig19

% figpoint2.eps  fig20

\newpage

\bibliographystyle{aasjournal}
\bibliography{smclastmosaics_ref}

\begin{thebibliography}{}
\expandafter\ifx\csname natexlab\endcsname\relax\def\natexlab#1{#1}\fi
\providecommand{\url}[1]{\href{#1}{#1}}
\providecommand{\dodoi}[1]{doi:~\href{http://doi.org/#1}{\nolinkurl{#1}}}
\providecommand{\doeprint}[1]{\href{http://ascl.net/#1}{\nolinkurl{http://ascl.net/#1}}}
\providecommand{\doarXiv}[1]{\href{https://arxiv.org/abs/#1}{\nolinkurl{https://arxiv.org/abs/#1}}}

\bibitem[{{Bolatto} {et~al.}(2007){Bolatto}, {Simon}, {Stanimirovi{\'c}}, {van
  Loon}, {Shah}, {Venn}, {Leroy}, {Sandstrom}, {Jackson}, {Israel}, {Li},
  {Staveley-Smith}, {Bot}, {Boulanger}, \& {Rubio}}]{bol07}
{Bolatto}, A.~D., {Simon}, J.~D., {Stanimirovi{\'c}}, S., {et~al.} 2007, \apj,
  655, 212, \dodoi{10.1086/509104}

\bibitem[{{Boyer} {et~al.}(2012){Boyer}, {Srinivasan}, {Riebel}, {McDonald},
  {van Loon}, {Clayton}, {Gordon}, {Meixner}, {Sargent}, \& {Sloan}}]{boy12}
{Boyer}, M.~L., {Srinivasan}, S., {Riebel}, D., {et~al.} 2012, \apj, 748, 40,
  \dodoi{10.1088/0004-637X/748/1/40}

\bibitem[{{Choudhury} {et~al.}(2020){Choudhury}, {de Grijs}, {Rubele}, {Bekki},
  {Cioni}, {Ivanov}, {van Loon}, {Niederhofer}, {Oliveira}, \&
  {Ripepi}}]{cho20}
{Choudhury}, S., {de Grijs}, R., {Rubele}, S., {et~al.} 2020, \mnras, 497,
  3746, \dodoi{10.1093/mnras/staa2140}

\bibitem[{{Cioni} {et~al.}(2011){Cioni}, {Clementini}, {Girardi}, {Guandalini},
  {Gullieuszik}, {Miszalski}, {Moretti}, {Ripepi}, {Rubele}, {Bagheri},
  {Bekki}, {Cross}, {de Blok}, {de Grijs}, {Emerson}, {Evans}, {Gibson},
  {Gonzales-Solares}, {Groenewegen}, {Irwin}, {Ivanov}, {Lewis}, {Marconi},
  {Marquette}, {Mastropietro}, {Moore}, {Napiwotzki}, {Naylor}, {Oliveira},
  {Read}, {Sutorius}, {van Loon}, {Wilkinson}, \& {Wood}}]{cio11}
{Cioni}, M.-R.~L., {Clementini}, G., {Girardi}, L., {et~al.} 2011, \aap, 527,
  A116, \dodoi{10.1051/0004-6361/201016137}

\bibitem[{{Cruz} {et~al.}(2003){Cruz}, {Reid}, {Liebert}, {Kirkpatrick}, \&
  {Lowrance}}]{cru03}
{Cruz}, K.~L., {Reid}, I.~N., {Liebert}, J., {Kirkpatrick}, J.~D., \&
  {Lowrance}, P.~J. 2003, \aj, 126, 2421, \dodoi{10.1086/378607}

\bibitem[{{Cutri} {et~al.}(2012){Cutri}, {Skrutskie}, {van Dyk}, {Beichman},
  {Carpenter}, {Chester}, {Cambresy}, {Evans}, {Fowler}, {Gizis}, {Howard},
  {Huchra}, {Jarrett}, {Kopan}, {Kirkpatrick}, {Light}, {Marsh}, {McCallon},
  {Schneider}, {Stiening}, {Sykes}, {Weinberg}, {Wheaton}, {Wheelock}, \&
  {Zacharias}}]{cutri12}
{Cutri}, R.~M., {Skrutskie}, M.~F., {van Dyk}, S., {et~al.} 2012, VizieR Online
  Data Catalog, II/281

\bibitem[{{de Grijs} \& {Bono}(2015)}]{deg15}
{de Grijs}, R., \& {Bono}, G. 2015, \aj, 149, 179,
  \dodoi{10.1088/0004-6256/149/6/179}

\bibitem[{{de Wit} {et~al.}(2003){de Wit}, {Beaulieu}, {Lamers}, {Lesquoy}, \&
  {Marquette}}]{wit03}
{de Wit}, W.~J., {Beaulieu}, J.~P., {Lamers}, H.~J.~G.~L.~M., {Lesquoy}, E., \&
  {Marquette}, J.~B. 2003, \aap, 410, 199, \dodoi{10.1051/0004-6361:20030948}

\bibitem[{{Fazio} {et~al.}(2004{\natexlab{a}}){Fazio}, {Hora}, {Allen},
  {Ashby}, {Barmby}, {Deutsch}, {Huang}, {Kleiner}, {Marengo}, {Megeath},
  {Melnick}, {Pahre}, {Patten}, {Polizotti}, {Smith}, {Taylor}, {Wang},
  {Willner}, {Hoffmann}, {Pipher}, {Forrest}, {McMurty}, {McCreight},
  {McKelvey}, {McMurray}, {Koch}, {Moseley}, {Arendt}, {Mentzell}, {Marx},
  {Losch}, {Mayman}, {Eichhorn}, {Krebs}, {Jhabvala}, {Gezari}, {Fixsen},
  {Flores}, {Shakoorzadeh}, {Jungo}, {Hakun}, {Workman}, {Karpati}, {Kichak},
  {Whitley}, {Mann}, {Tollestrup}, {Eisenhardt}, {Stern}, {Gorjian},
  {Bhattacharya}, {Carey}, {Nelson}, {Glaccum}, {Lacy}, {Lowrance}, {Laine},
  {Reach}, {Stauffer}, {Surace}, {Wilson}, {Wright}, {Hoffman}, {Domingo}, \&
  {Cohen}}]{faz04}
{Fazio}, G.~G., {Hora}, J.~L., {Allen}, L.~E., {et~al.} 2004{\natexlab{a}},
  \apjs, 154, 10, \dodoi{10.1086/422843}

\bibitem[{{Fazio} {et~al.}(2004{\natexlab{b}}){Fazio}, {Hora}, {Allen},
  {Ashby}, {Barmby}, {Deutsch}, {Huang}, {Kleiner}, {Marengo}, {Megeath},
  {Melnick}, {Pahre}, {Patten}, {Polizotti}, {Smith}, {Taylor}, {Wang},
  {Willner}, {Hoffmann}, {Pipher}, {Forrest}, {McMurty}, {McCreight},
  {McKelvey}, {McMurray}, {Koch}, {Moseley}, {Arendt}, {Mentzell}, {Marx},
  {Losch}, {Mayman}, {Eichhorn}, {Krebs}, {Jhabvala}, {Gezari}, {Fixsen},
  {Flores}, {Shakoorzadeh}, {Jungo}, {Hakun}, {Workman}, {Karpati}, {Kichak},
  {Whitley}, {Mann}, {Tollestrup}, {Eisenhardt}, {Stern}, {Gorjian},
  {Bhattacharya}, {Carey}, {Nelson}, {Glaccum}, {Lacy}, {Lowrance}, {Laine},
  {Reach}, {Stauffer}, {Surace}, {Wilson}, {Wright}, {Hoffman}, {Domingo}, \&
  {Cohen}}]{irac04}
---. 2004{\natexlab{b}}, \apjs, 154, 10, \dodoi{10.1086/422843}

\bibitem[{{Gordon} {et~al.}(2011{\natexlab{a}}){Gordon}, {Meixner}, {Meade},
  {Whitney}, {Engelbracht}, {Bot}, {Boyer}, {Lawton}, {Sewi{\l}o}, {Babler},
  {Bernard}, {Bracker}, {Block}, {Blum}, {Bolatto}, {Bonanos}, {Harris},
  {Hora}, {Indebetouw}, {Misselt}, {Reach}, {Shiao}, {Tielens}, {Carlson},
  {Churchwell}, {Clayton}, {Chen}, {Cohen}, {Fukui}, {Gorjian}, {Hony},
  {Israel}, {Kawamura}, {Kemper}, {Leroy}, {Li}, {Madden}, {Marble},
  {McDonald}, {Mizuno}, {Mizuno}, {Muller}, {Oliveira}, {Olsen}, {Onishi},
  {Paladini}, {Paradis}, {Points}, {Robitaille}, {Rubin}, {Sandstrom}, {Sato},
  {Shibai}, {Simon}, {Smith}, {Srinivasan}, {Vijh}, {Van Dyk}, {van Loon}, \&
  {Zaritsky}}]{gor11}
{Gordon}, K.~D., {Meixner}, M., {Meade}, M.~R., {et~al.} 2011{\natexlab{a}},
  \aj, 142, 102, \dodoi{10.1088/0004-6256/142/4/102}

\bibitem[{{Gordon} {et~al.}(2011{\natexlab{b}}){Gordon}, {Meixner}, {Meade},
  {Whitney}, {Engelbracht}, {Bot}, {Boyer}, {Lawton}, {Sewi{\l}o}, {Babler},
  {Bernard}, {Bracker}, {Block}, {Blum}, {Bolatto}, {Bonanos}, {Harris},
  {Hora}, {Indebetouw}, {Misselt}, {Reach}, {Shiao}, {Tielens}, {Carlson},
  {Churchwell}, {Clayton}, {Chen}, {Cohen}, {Fukui}, {Gorjian}, {Hony},
  {Israel}, {Kawamura}, {Kemper}, {Leroy}, {Li}, {Madden}, {Marble},
  {McDonald}, {Mizuno}, {Mizuno}, {Muller}, {Oliveira}, {Olsen}, {Onishi},
  {Paladini}, {Paradis}, {Points}, {Robitaille}, {Rubin}, {Sandstrom}, {Sato},
  {Shibai}, {Simon}, {Smith}, {Srinivasan}, {Vijh}, {Van Dyk}, {van Loon}, \&
  {Zaritsky}}]{sagesmc11}
---. 2011{\natexlab{b}}, \aj, 142, 102, \dodoi{10.1088/0004-6256/142/4/102}

\bibitem[{{Graczyk} {et~al.}(2020){Graczyk}, {Pietrzy{\'n}ski}, {Thompson},
  {Gieren}, {Zgirski}, {Villanova}, {G{\'o}rski}, {Wielg{\'o}rski},
  {Karczmarek}, {Narloch}, {Pilecki}, {Taormina}, {Smolec}, {Suchomska},
  {Gallenne}, {Nardetto}, {Storm}, {Kudritzki}, {Ka{\l}uszy{\'n}ski}, \&
  {Pych}}]{gra20}
{Graczyk}, D., {Pietrzy{\'n}ski}, G., {Thompson}, I.~B., {et~al.} 2020, \apj,
  904, 13, \dodoi{10.3847/1538-4357/abbb2b}

\bibitem[{{Groenewegen} \& {Sloan}(2018)}]{gro18}
{Groenewegen}, M.~A.~T., \& {Sloan}, G.~C. 2018, \aap, 609, A114,
  \dodoi{10.1051/0004-6361/201731089}

\bibitem[{{Hora} {et~al.}(2006){Hora}, {Patten}, {Fazio}, \&
  {Glaccum}}]{horaea06}
{Hora}, J.~L., {Patten}, B.~M., {Fazio}, G.~G., \& {Glaccum}, W.~J. 2006, in
  \procspie, Vol. 6276, Society of Photo-Optical Instrumentation Engineers
  (SPIE) Conference Series, 62760J, \dodoi{10.1117/12.672017}

\bibitem[{{Hora} {et~al.}(2004){Hora}, {Fazio}, {Allen}, {Ashby}, {Barmby},
  {Deutsch}, {Huang}, {Marengo}, {Megeath}, {Melnick}, {Pahre}, {Patten},
  {Smith}, {Wang}, {Willner}, {Hoffmann}, {Pipher}, {Forrest}, {McMurtry},
  {McCreight}, {McKelvey}, {McMurray}, {Moseley}, {Arendt}, {Mentzell}, {Marx},
  {Fixsen}, {Tollestrup}, {Eisenhardt}, {Stern}, {Gorjian}, {Bhattacharya},
  {Carey}, {Glaccum}, {Lacy}, {Lowrance}, {Laine}, {Nelson}, {Reach},
  {Stauffer}, {Surace}, {Wilson}, \& {Wright}}]{horaea04}
{Hora}, J.~L., {Fazio}, G.~G., {Allen}, L.~E., {et~al.} 2004, in \procspie,
  Vol. 5487, Optical, Infrared, and Millimeter Space Telescopes, ed. J.~C.
  {Mather}, 77--92, \dodoi{10.1117/12.550744}

\bibitem[{{Hora} {et~al.}(2008){Hora}, {Carey}, {Surace}, {Marengo},
  {Lowrance}, {Glaccum}, {Lacy}, {Reach}, {Hoffmann}, {Barmby}, {Willner},
  {Fazio}, {Megeath}, {Allen}, {Bhattacharya}, \& {Quijada}}]{horaea08irac}
{Hora}, J.~L., {Carey}, S., {Surace}, J., {et~al.} 2008, \pasp, 120, 1233,
  \dodoi{10.1086/593217}

\bibitem[{{IRAC Instrument and Instrument Support Teams}(2021)}]{irachb21}
{IRAC Instrument and Instrument Support Teams}. 2021, {IRAC Instrument
  Handbook}, Tech. rep., {Infrared Science Archive}, \dodoi{10.26131/irsa486}

\bibitem[{{Ita} {et~al.}(2018){Ita}, {Matsunaga}, {Tanab{\'e}}, {Nakada},
  {Kato}, {Nagayama}, {Nagashima}, {Kurita}, {Nakajima}, {Whitelock},
  {Menzies}, {Feast}, {Nagata}, {Tamura}, \& {Nakaya}}]{ita18}
{Ita}, Y., {Matsunaga}, N., {Tanab{\'e}}, T., {et~al.} 2018, \mnras, 481, 4206,
  \dodoi{10.1093/mnras/sty2539}

\bibitem[{{Joy}(1945)}]{joy45}
{Joy}, A.~H. 1945, \apj, 102, 168, \dodoi{10.1086/144749}

\bibitem[{{Kirkpatrick} {et~al.}(2011){Kirkpatrick}, {Cushing}, {Gelino},
  {Griffith}, {Skrutskie}, {Marsh}, {Wright}, {Mainzer}, {Eisenhardt},
  {McLean}, {Thompson}, {Bauer}, {Benford}, {Bridge}, {Lake}, {Petty},
  {Stanford}, {Tsai}, {Bailey}, {Beichman}, {Bloom}, {Bochanski}, {Burgasser},
  {Capak}, {Cruz}, {Hinz}, {Kartaltepe}, {Knox}, {Manohar}, {Masters},
  {Morales-Calder{\'o}n}, {Prato}, {Rodigas}, {Salvato}, {Schurr}, {Scoville},
  {Simcoe}, {Stapelfeldt}, {Stern}, {Stock}, \& {Vacca}}]{kir11}
{Kirkpatrick}, J.~D., {Cushing}, M.~C., {Gelino}, C.~R., {et~al.} 2011, \apjs,
  197, 19, \dodoi{10.1088/0067-0049/197/2/19}

\bibitem[{{Kirkpatrick} {et~al.}(2012){Kirkpatrick}, {Gelino}, {Cushing},
  {Mace}, {Griffith}, {Skrutskie}, {Marsh}, {Wright}, {Eisenhardt}, {McLean},
  {Mainzer}, {Burgasser}, {Tinney}, {Parker}, \& {Salter}}]{kir12}
{Kirkpatrick}, J.~D., {Gelino}, C.~R., {Cushing}, M.~C., {et~al.} 2012, \apj,
  753, 156, \dodoi{10.1088/0004-637X/753/2/156}

\bibitem[{{Luck} {et~al.}(1998){Luck}, {Moffett}, {Barnes}, \&
  {Gieren}}]{luc98}
{Luck}, R.~E., {Moffett}, T.~J., {Barnes}, Thomas~G., I., \& {Gieren}, W.~P.
  1998, \aj, 115, 605, \dodoi{10.1086/300227}

\bibitem[{{Mainzer} {et~al.}(2014){Mainzer}, {Bauer}, {Cutri}, {Grav},
  {Masiero}, {Beck}, {Clarkson}, {Conrow}, {Dailey}, {Eisenhardt}, {Fabinsky},
  {Fajardo-Acosta}, {Gelino}, {Grillmair}, {Heinrichsen}, {Kendall},
  {Kirkpatrick}, {Liu}, {Masci}, {McCallon}, {Nugent}, {Papin}, {Rice},
  {Royer}, {Ryan}, {Sevilla}, {Sonnett}, {Stevenson}, {Thompson}, {Wheelock},
  {Wiemer}, {Wittman}, {Wright}, \& {Yan}}]{mai14}
{Mainzer}, A., {Bauer}, J., {Cutri}, R.~M., {et~al.} 2014, \apj, 792, 30,
  \dodoi{10.1088/0004-637X/792/1/30}

\bibitem[{{Makovoz} \& {Khan}(2005)}]{mopex05}
{Makovoz}, D., \& {Khan}, I. 2005, in Astronomical Society of the Pacific
  Conference Series, Vol. 347, Astronomical Data Analysis Software and Systems
  XIV, ed. P.~{Shopbell}, M.~{Britton}, \& R.~{Ebert}, 81

\bibitem[{{Megeath} {et~al.}(2012){Megeath}, {Gutermuth}, {Muzerolle},
  {Kryukova}, {Flaherty}, {Hora}, {Allen}, {Hartmann}, {Myers}, {Pipher},
  {Stauffer}, {Young}, \& {Fazio}}]{meg12}
{Megeath}, S.~T., {Gutermuth}, R., {Muzerolle}, J., {et~al.} 2012, \aj, 144,
  192, \dodoi{10.1088/0004-6256/144/6/192}

\bibitem[{{Meixner} {et~al.}(2006){Meixner}, {Gordon}, {Indebetouw}, {Hora},
  {Whitney}, {Blum}, {Reach}, {Bernard}, {Meade}, {Babler}, {Engelbracht},
  {For}, {Misselt}, {Vijh}, {Leitherer}, {Cohen}, {Churchwell}, {Boulanger},
  {Frogel}, {Fukui}, {Gallagher}, {Gorjian}, {Harris}, {Kelly}, {Kawamura},
  {Kim}, {Latter}, {Madden}, {Markwick-Kemper}, {Mizuno}, {Mizuno}, {Mould},
  {Nota}, {Oey}, {Olsen}, {Onishi}, {Paladini}, {Panagia}, {Perez-Gonzalez},
  {Shibai}, {Sato}, {Smith}, {Staveley-Smith}, {Tielens}, {Ueta}, {van Dyk},
  {Volk}, {Werner}, \& {Zaritsky}}]{mei06}
{Meixner}, M., {Gordon}, K.~D., {Indebetouw}, R., {et~al.} 2006, \aj, 132,
  2268, \dodoi{10.1086/508185}

\bibitem[{{Mizuno}(2008)}]{mizuno08}
{Mizuno}, D.~R. 2008, AFRL$\_$BCD$\_$OVERLAP, Contributed software, Spitzer
  Science Center, irsa.ipac.caltech.edu/data/SPITZER/docs/
  dataanalysistools/tools/contributed/irac/ afrlbcdoverlap/.
\newblock
  \url{http://irsa.ipac.caltech.edu/data/SPITZER/docs/dataanalysistools/tools/contributed/irac/afrlbcdoverlap/}

\bibitem[{{Mizuno} {et~al.}(2008){Mizuno}, {Carey}, {Noriega-Crespo},
  {Paladini}, {Padgett}, {Shenoy}, {Kuchar}, {Kraemer}, \&
  {Price}}]{mizunoea08b}
{Mizuno}, D.~R., {Carey}, S.~J., {Noriega-Crespo}, A., {et~al.} 2008, \pasp,
  120, 1028, \dodoi{10.1086/591809}

\bibitem[{{Nidever} {et~al.}(2013){Nidever}, {Monachesi}, {Bell}, {Majewski},
  {Mu{\~n}oz}, \& {Beaton}}]{nid13}
{Nidever}, D.~L., {Monachesi}, A., {Bell}, E.~F., {et~al.} 2013, \apj, 779,
  145, \dodoi{10.1088/0004-637X/779/2/145}

\bibitem[{{Rice} {et~al.}(2015){Rice}, {Reipurth}, {Wolk}, {Vaz}, \&
  {Cross}}]{ric15}
{Rice}, T.~S., {Reipurth}, B., {Wolk}, S.~J., {Vaz}, L.~P., \& {Cross},
  N.~J.~G. 2015, \aj, 150, 132, \dodoi{10.1088/0004-6256/150/4/132}

\bibitem[{{Rice} {et~al.}(2012){Rice}, {Wolk}, \& {Aspin}}]{ric12}
{Rice}, T.~S., {Wolk}, S.~J., \& {Aspin}, C. 2012, \apj, 755, 65,
  \dodoi{10.1088/0004-637X/755/1/65}

\bibitem[{{Riebel} {et~al.}(2015){Riebel}, {Boyer}, {Srinivasan}, {Whitelock },
  {Meixner}, {Babler}, {Feast}, {Groenewegen}, {Ita}, {Meade}, {Shiao}, \&
  {Whitney}}]{rie15}
{Riebel}, D., {Boyer}, M.~L., {Srinivasan}, S., {et~al.} 2015, \apj, 807, 1,
  \dodoi{10.1088/0004-637X/807/1/1}

\bibitem[{{Rieke} {et~al.}(2004){Rieke}, {Young}, {Engelbracht}, {Kelly},
  {Low}, {Haller}, {Beeman}, {Gordon}, {Stansberry}, {Misselt}, {Cadien},
  {Morrison}, {Rivlis}, {Latter}, {Noriega-Crespo}, {Padgett}, {Stapelfeldt},
  {Hines}, {Egami}, {Muzerolle}, {Alonso-Herrero}, {Blaylock}, {Dole}, {Hinz},
  {Le Floc'h}, {Papovich}, {P{\'e}rez-Gonz{\'a}lez}, {Smith}, {Su}, {Bennett},
  {Frayer}, {Henderson}, {Lu}, {Masci}, {Pesenson}, {Rebull}, {Rho}, {Keene},
  {Stolovy}, {Wachter}, {Wheaton}, {Werner}, \& {Richards}}]{rie04}
{Rieke}, G.~H., {Young}, E.~T., {Engelbracht}, C.~W., {et~al.} 2004, \apjs,
  154, 25, \dodoi{10.1086/422717}

\bibitem[{{Rubele} {et~al.}(2018){Rubele}, {Pastorelli}, {Girardi}, {Cioni},
  {Zaggia}, {Marigo}, {Bekki}, {Bressan}, {Clementini}, {de Grijs}, {Emerson},
  {Groenewegen}, {Ivanov}, {Muraveva}, {Nanni}, {Oliveira}, {Ripepi}, {Sun}, \&
  {van Loon}}]{rub18}
{Rubele}, S., {Pastorelli}, G., {Girardi}, L., {et~al.} 2018, \mnras, 478,
  5017, \dodoi{10.1093/mnras/sty1279}

\bibitem[{{Scowcroft} {et~al.}(2016){Scowcroft}, {Freedman}, {Madore},
  {Monson}, {Persson}, {Rich}, {Seibert}, \& {Rigby}}]{sco16}
{Scowcroft}, V., {Freedman}, W.~L., {Madore}, B.~F., {et~al.} 2016, \apj, 816,
  49, \dodoi{10.3847/0004-637X/816/2/49}

\bibitem[{{Skrutskie} {et~al.}(2006){Skrutskie}, {Cutri}, {Stiening},
  {Weinberg}, {Schneider}, {Carpenter}, {Beichman}, {Capps}, {Chester},
  {Elias}, {Huchra}, {Liebert}, {Lonsdale}, {Monet}, {Price}, {Seitzer},
  {Jarrett}, {Kirkpatrick}, {Gizis}, {Howard}, {Evans}, {Fowler}, {Fullmer},
  {Hurt}, {Light}, {Kopan}, {Marsh}, {McCallon}, {Tam}, {Van Dyk}, \&
  {Wheelock}}]{skrut06}
{Skrutskie}, M.~F., {Cutri}, R.~M., {Stiening}, R., {et~al.} 2006, \aj, 131,
  1163, \dodoi{10.1086/498708}

\bibitem[{{Sloan} {et~al.}(2016){Sloan}, {Kraemer}, {McDonald}, {Groenewegen},
  {Wood}, {Zijlstra}, {Lagadec}, {Boyer}, {Kemper}, {Matsuura}, {Sahai},
  {Sargent}, {Srinivasan}, {van Loon}, \& {Volk}}]{slo16}
{Sloan}, G.~C., {Kraemer}, K.~E., {McDonald}, I., {et~al.} 2016, \apj, 826, 44,
  \dodoi{10.3847/0004-637X/826/1/44}

\bibitem[{{Soszy{\'n}ski} {et~al.}(2011){Soszy{\'n}ski}, {Udalski},
  {Szyma{\'n}ski}, {Kubiak}, {Pietrzy{\'n}ski}, {Wyrzykowski}, {Ulaczyk},
  {Poleski}, {Koz{\l}owski}, \& {Pietrukowicz}}]{sos11}
{Soszy{\'n}ski}, I., {Udalski}, A., {Szyma{\'n}ski}, M.~K., {et~al.} 2011,
  \actaa, 61, 217.
\newblock \doarXiv{1109.1143}

\bibitem[{{Srinivasan} {et~al.}(2016){Srinivasan}, {Boyer}, {Kemper},
  {Meixner}, {Sargent}, \& {Riebel}}]{sri16}
{Srinivasan}, S., {Boyer}, M.~L., {Kemper}, F., {et~al.} 2016, \mnras, 457,
  2814, \dodoi{10.1093/mnras/stw155}

\bibitem[{{Subramanian} {et~al.}(2017){Subramanian}, {Rubele}, {Sun},
  {Girardi}, {de Grijs}, {van Loon}, {Cioni}, {Piatti}, {Bekki}, {Emerson},
  {Ivanov}, {Kerber}, {Marconi}, {Ripepi}, \& {Tatton}}]{sub17}
{Subramanian}, S., {Rubele}, S., {Sun}, N.-C., {et~al.} 2017, \mnras, 467,
  2980, \dodoi{10.1093/mnras/stx205}

\bibitem[{{Udalski} {et~al.}(1997){Udalski}, {Kubiak}, \& {Szymanski}}]{uda97}
{Udalski}, A., {Kubiak}, M., \& {Szymanski}, M. 1997, \actaa, 47, 319.
\newblock \doarXiv{astro-ph/9710091}

\bibitem[{{Udalski} {et~al.}(2008){Udalski}, {Soszy{\'n}ski}, {Szyma{\'n}ski},
  {Kubiak}, {Pietrzy{\'n}ski}, {Wyrzykowski}, {Szewczyk}, {Ulaczyk}, \&
  {Poleski}}]{uda08}
{Udalski}, A., {Soszy{\'n}ski}, I., {Szyma{\'n}ski}, M.~K., {et~al.} 2008,
  \actaa, 58, 329.
\newblock \doarXiv{0901.4632}

\bibitem[{{Whitelock} {et~al.}(1994){Whitelock}, {Menzies}, {Feast}, {Marang},
  {Carter}, {Roberts}, {Catchpole}, \& {Chapman}}]{whi94}
{Whitelock}, P., {Menzies}, J., {Feast}, M., {et~al.} 1994, \mnras, 267, 711,
  \dodoi{10.1093/mnras/267.3.711}

\bibitem[{{Wright} {et~al.}(2010){Wright}, {Eisenhardt}, {Mainzer}, {Ressler},
  {Cutri}, {Jarrett}, {Kirkpatrick}, {Padgett}, {McMillan}, {Skrutskie},
  {Stanford}, {Cohen}, {Walker}, {Mather}, {Leisawitz}, {Gautier}, {McLean},
  {Benford}, {Lonsdale}, {Blain}, {Mendez}, {Irace}, {Duval}, {Liu}, {Royer},
  {Heinrichsen}, {Howard}, {Shannon}, {Kendall}, {Walsh}, {Larsen}, {Cardon},
  {Schick}, {Schwalm}, {Abid}, {Fabinsky}, {Naes}, \& {Tsai}}]{wri10}
{Wright}, E.~L., {Eisenhardt}, P.~R.~M., {Mainzer}, A.~K., {et~al.} 2010, \aj,
  140, 1868, \dodoi{10.1088/0004-6256/140/6/1868}

\bibitem[{{Yanchulova Merica-Jones} {et~al.}(2021){Yanchulova Merica-Jones},
  {Sandstrom}, {Johnson}, {Dolphin}, {Dalcanton}, {Gordon}, {Roman-Duval},
  {Weisz}, \& {Williams}}]{yan21}
{Yanchulova Merica-Jones}, P., {Sandstrom}, K.~M., {Johnson}, L.~C., {et~al.}
  2021, \apj, 907, 50, \dodoi{10.3847/1538-4357/abc48b}

\end{thebibliography}

\end{document}